\documentstyle[11pt,aaspp4,psfig]{article}
\tighten 
\slugcomment{Accepted for publication in {\it The Astrophysical Journal}}
\def\spose#1{\hbox to 0pt{#1\hss}}

\def\kms{\ifmmode {\rm\,km\,s^{-1}}\else ${\rm\,km\,s^{-1}}$\fi}

\def\kmsmpc{\ifmmode {\rm\,km\,s^{-1}\,Mpc^{-1}}\else ${\rm\,km\,s^{-1}\,Mpc^{-1}}$\fi}

\def\ergps{\ifmmode {\rm\,erg\,s^{-1}}\else ${\rm\,erg\,s^{-1}}$\fi}
\def\ergpscm2{\ifmmode {\rm\,erg\,s^{-1}\,cm^{-2}}\else
    ${\rm\,erg\,s^{-1}\,cm^{-2}}$\fi}

\def\deg{\ifmmode {^{\circ}}\else {$^\circ$}\fi}
\def\degr{\ifmmode {^{\circ}}\else {$^\circ$}\fi}
\def\degs{\ifmmode {^{\circ}}\else {$^\circ$}\fi}
\def\degspt{$^\circ_\cdot$}
\def\etal{{\it et al.~}}

\def\h3Mpc{h^{-3}{\rm Mpc}^3}

\def\arcsec{\ifmmode {^{\prime\prime}}\else $^{\prime\prime}$\fi}
\def\asec{\ifmmode {^{\prime\prime}}\else $^{\prime\prime}$\fi}
\def\arcmin{\ifmmode {^{\prime}}\else $^{\prime}$\fi}
\def\amin{\ifmmode {^{\prime}}\else $^{\prime}$\fi}

\def\secper{\ifmmode \rlap.{^{s}}\else $\rlap{.}{^{s}} $\fi}
\def\minper{\ifmmode \rlap.{^{m}}\else $\rlap{.}{^m} $\fi}
\def\secspt{\ifmmode \rlap.{^{\prime\prime}}\else
    $\rlap.{^{\prime\prime}}$\fi}
\def\arcsper{\ifmmode \rlap.{^{\prime\prime}}\else
    $\rlap.{^{\prime\prime}}$\fi}
\def\minspt{\ifmmode \rlap.{^{\prime}}\else
    $\rlap.{^{\prime}}$\fi}
\def\arcmper{\ifmmode \rlap.{^{\prime}}\else
    $\rlap.{^{\prime}}$\fi}
\def\spose#1{\hbox to 0pt{#1\hss}}
\def\simlt{\mathrel{\spose{\lower 3pt\hbox{$\mathchar"218$}}
     \raise 2.0pt\hbox{$\mathchar"13C$}}}
\def\simgt{\mathrel{\spose{\lower 3pt\hbox{$\mathchar"218$}}
     \raise 2.0pt\hbox{$\mathchar"13E$}}}

\def\refindent{\par\noindent\parskip=2pt\hangindent=3pc\hangafter=1 }

\def\aa{{A\&A}}

\def\aasup{{A\&AS}}
\def\aj{{AJ}}
\def\apj{{ApJ}}

\def\apjs{{ApJS}}

\def\mnras{{MNRAS}}
\def\nature{{Nature}}

\def\pasp{{PASP}}

\def\ref#1;#2;#3;#4 {\refindent{#1,} {#2}, #3, #4}

\def\book#1;#2;#3 {\refindent{#1, }{in {\it{#2},} }{#3}}

\def\U300{\ifmmode{U_{300}}\else{$U_{300}$}\fi}
\def\B450{\ifmmode{B_{450}}\else{$B_{450}$}\fi}
\def\V606{\ifmmode{V_{606}}\else{$V_{606}$}\fi}
\def\I814{\ifmmode{I_{814}}\else{$I_{814}$}\fi}
\def\J110{\ifmmode{J_{110}}\else{$J_{110}$}\fi}
\def\H160{\ifmmode{H_{160}}\else{$H_{160}$}\fi}
\def\jdrop{HDFN--JD1}

\def\jd1{JD1}

\begin{document}

\title{The Unusual Infrared Object HDF--N~J123656.3+621322\altaffilmark{1,3}}
\author{\sc Mark Dickinson\altaffilmark{2} and Christopher Hanley}
\affil{Space Telescope Science Institute, 3700 San Martin Dr.,
	Baltimore MD 21218}

\author{\sc Richard Elston\altaffilmark{2}}
\affil{Department of Astronomy, University of Florida,
	Gainesville, FL, 32611}

\author{\sc Peter R.\ Eisenhardt\altaffilmark{2}}
\affil{MS 169-327, Jet Propulsion Laboratory,
	California Institute of Technology,
	4800 Oak Grove Drive, Pasadena, CA, 91109}

\author{\sc S.\ A.\ Stanford\altaffilmark{2,4}}
\affil{Physics Department, University of California, Davis, CA 95616}

%\affil{University of California at Davis, and the Institute of Geophysics and 
%	Planetary Physics, Lawrence Livermore National Laboratories, Livermore, 
%	CA, 94550}

\author{\sc Kurt L.\ Adelberger, Alice Shapley, and Charles C.\ Steidel}
\affil{Palomar Observatory, Caltech 105--24, Pasadena, CA 91125}

\author{\sc Casey Papovich and Alexander S.\ Szalay}
\affil{Department of Physics and Astronomy,
	The Johns Hopkins University, 3400 N.\ Charles St., Baltimore MD 21218}

\author{\sc Matthew A.\ Bershady\altaffilmark{2} and Christopher J.\ Conselice}
\affil{Department of Astronomy, University of Wisconsin,
	475 N.\ Charter St., Madison, WI 53706}

\author{\sc Henry C.\ Ferguson and Andrew S.\ Fruchter}
\affil{Space Telescope Science Institute, 3700 San Martin Dr.,
	Baltimore MD 21218}

\altaffiltext{1}{Based on observations with the NASA/ESA
Hubble Space Telescope, obtained at the Space Telescope Science Institute, 
which is operated by the Association of Universities for Research in 
Astronomy, Inc., under NASA contract NAS5-26555.}
 
\altaffiltext{2}{Visiting Astronomer, Kitt Peak National Observatory, 
National Optical Astronomy Observatories, which is operated by the Association 
of Universities for Research in Astronomy, Inc.\ (AURA) under cooperative agreement 
with the National Science Foundation.}

\altaffiltext{3}{Partially based on data obtained at the W.M.~Keck Observatory,
operated as a scientific partnership among the California Institute of Technology, 
the University of California, and NASA, and made possible by the 
generous financial support of the W.M.~Keck Foundation.}

\altaffiltext{4}{Institute of Geophysics and Planetary Physics, Lawrence Livermore 
National Laboratories, Livermore, CA, 94550}

\clearpage

\begin{abstract}
We describe an object in the Hubble Deep Field North with very unusual 
near--infrared properties.  It is readily visible in {\it Hubble Space 
Telescope} NICMOS images at $1.6\mu$m and from the ground at 2.2$\mu$m,
but is undetected (with signal--to--noise~$\simlt 2$) in very deep WFPC2
and NICMOS data from 0.3 to 1.1$\mu$m.  The $f_{\nu}$ flux density drops
by a factor $\simgt 8.3$ (97.7\% confidence) from 1.6 to 1.1$\mu$m.  
The object is compact but may be slightly resolved in the NICMOS 1.6$\mu$m 
image.  In a low--resolution, near--infrared spectrogram, we find a 
possible emission line at $1.643\mu$m, but a reobservation at higher 
spectral resolution failed to confirm the line, leaving its reality 
in doubt.  We consider various hypotheses for the nature of this object.
Its colors are unlike those of known galactic stars, except perhaps the 
most extreme carbon stars or Mira variables with thick circumstellar 
dust shells.  It does not appear to be possible to explain its spectral 
energy distribution as that of a normal galaxy at any redshift without
additional opacity from either dust or intergalactic neutral hydrogen.
The colors can be matched by those of a dusty galaxy at $z \simgt 2$, 
by a maximally old elliptical galaxy at $z \simgt 3$ (perhaps with 
some additional reddening), or by an object at $z \simgt 10$ whose 
optical and 1.1$\mu$m light have been suppressed by the intergalactic
medium.   Under the latter hypothesis, if the luminosity results from
stars and not an AGN, the object would resemble a classical, unobscured 
protogalaxy, with a star formation rate $\simgt 100 M_\odot$~yr$^{-1}$.
Such UV--bright objects are evidently rare at $2 < z < 12.5$, however, 
with a space density several hundred times lower than that of 
present--day $L^\ast$ galaxies.
\end{abstract}


\clearpage

\section{Introduction}

In recent years, astronomers have extensively developed the art of selecting 
interesting, high redshift objects on the basis of their broad band colors.
Color selection has long been used to separate distant quasar candidates 
from the multitude of foreground stars either via their UV excess 
(e.g., Veron 1983) or by color criteria based on the passage of the 1216\AA\ 
Lyman~$\alpha$ forest and 912\AA\ Lyman limit discontinuities through broad 
passbands (e.g., Warren \etal 1987).  Guhathakurta, Tyson, \& Majewski (1990) 
applied the latter technique to search for high redshift galaxies, and the 
method was brought to fruition by Steidel \etal (1996a,b, 1999) who have 
successfully identified and spectroscopically confirmed nearly 1000 galaxies 
at $2 < z < 4.5$ using this approach.   Other interesting classes of 
objects have been selected and studied on the basis of having extremely 
red optical--to--infrared colors (e.g., Elston, Rieke, \& Rieke 1988, 1989; 
Hu \& Ridgway 1994; Graham \& Dey 1996;  Thompson \etal 1999a).  Some of these 
turn out to be early type galaxies at high redshift, while others are both distant 
and dust--obscured.  In general, estimating galaxy redshifts from broad 
band colors is now a popular industry, and much progress has been made 
in applying such techniques at intermediate redshifts (e.g., Brunner, 
Connolly \& Szalay 1999) and for identifying candidates for extremely 
distant galaxies (Lanzetta, Yahil, \& Fern\'andez--Soto 1996).  Indeed, 
some of the most distant galaxies now known have been found in this way
(e.g., HDF 4-473, Weymann \etal 1998; HDF 3-951, 
Spinrad \etal 1998\footnote{We will occasionally use catalog numbers 
from Williams \etal 1996 to identify HDF galaxies in this paper.}).

The Hubble Deep Fields (North and South, or HDF--N and HDF--S, 
Williams \etal 1996 and 1999) have been valuable data sets for exploring 
such techniques because of the extremely deep, multiwavelength data which 
are available.   The HDF--N has recently been the subject of two 
near--infrared surveys with NICMOS (Thompson \etal 1998) on board the 
{\it Hubble Space Telescope} (HST):  a deep image of a $\sim 1$~arcmin$^2$ 
sub--region (Thompson \etal 1999b), and a shallower map of the entire
field (Dickinson \etal 1999; see also Dickinson 1999).  These data offer 
new opportunities for identifying interesting objects on the basis of 
their colors.  Here we describe an object with perhaps the most unusual 
colors in the HDF--N, which is significantly detected only at 
$\lambda \geq 1.6\mu$m.  The object was first noted by Lanzetta, 
Yahil, \& Fern\'andez--Soto (1998), who used the ground--based 
infrared images which we obtained at Kitt Peak National Observatory 
(cf.\ Dickinson 1998) to identify candidate $K$--band sources 
without optical counterparts.  We discuss our imaging and 
spectroscopic observations of this object, and consider various 
interpretations of its nature.

\section{Imaging and Photometry}

We observed the HDF--N with NICMOS between 
UT 1998 June 13 and June 23, during the second refocus campaign 
when the HST secondary mirror was moved to ensure optimal focus 
for NICMOS Camera 3.   The observations and data reduction will be 
described in detail elsewhere (Dickinson \etal 1999);  
the immediately relevant aspects are summarized here.  The complete 
HDF--N was mosaiced with 8 sub--fields, each imaged during three 
separate visits.  During each visit, exposures were taken through both 
the F110W (1.1$\mu$m) and F160W (1.6$\mu$m) filters.  (Henceforth we 
will refer to the six WFPC2 and NICMOS HDF bandpasses as \U300, \B450, 
\V606, \I814, \J110 and \H160.)  Each section of the mosaic was 
dithered through 9 independent positions, with a net exposure time 
of 12600s per filter, except in a few cases where telescope tracking 
was lost due to HST Fine Guidance Sensor failures.  The region of 
interest for this paper did not lose any exposure time.  The data 
were processed using a combination of STScI pipeline routines and 
custom software, and were combined into a single mosaic, accurately 
registered to the HDF--N WFPC2 images, using the ``drizzling'' method 
of Fruchter \& Hook (1999).  The NICMOS images have a point spread 
function (PSF) with FWHM~$\approx$~0\secspt22, primarily limited by 
the pixel scale (0\secspt2) of Camera~3.  Sensitivity varies over the 
field of view due to variations in NICMOS quantum efficiency and 
exposure time, but on average the images have a signal--to--noise ratio
$S/N \approx 10$ within an $0\secspt7$ diameter aperture at $AB \approx 26.1$ 
for both the \J110 and \H160 filters.\footnote{Unless otherwise stated, 
we use AB magnitudes throughout this paper, defined as 
$AB = 31.4 - 2.5\log\langle f_\nu \rangle$, where $\langle f_\nu \rangle$
is the flux density in nJy averaged over the filter bandpass.}  
In order to ensure properly matched photometry between the optical 
and infrared images, the WFPC2 data were convolved to match the 
NICMOS PSF.  Photometric catalogs were constructed using SExtractor 
(Bertin \& Arnouts 1996), by detecting objects in the NICMOS images
and measuring fluxes through matched apertures in all bands.

We noticed the object which we will call HDF--N~J123656.3+621322, or 
\jdrop\ for brevity, during an initial visual inspection of the NICMOS 
data for objects with unusual colors.  It is prominent at 1.6$\mu$m, but 
apparently invisible through all other HST filters, including \J110 
(Figure~1).  Figure~2 shows a $\J110 - \H160$ vs.\ \H160 color--magnitude 
diagram highlighting the object's extreme and unusual color.  
In $\J110 - \H160$, \jdrop\ is by far the reddest among the $\sim 1700$ 
objects detected in our NICMOS survey.  

Careful inspection of the NICMOS data convinces us that this is a real
astronomical source, not an artifact, and that there is no evidence 
that it is transient or variable.   It is visible in each of the nine 
individual, dithered \H160 exposures, which were taken during telescope 
visits on UT 1998 June 16, 20 and 22--23,  but is not detected in 
any of the corresponding \J110 exposures taken during those same visits.  
We know of no NICMOS anomaly or optics ghost which could produce
an artifact resembling what we observe.\footnote{One NICMOS anomaly, 
the ``Mr. Staypuft Effect'' (cf.\ Skinner \etal 1998), produces electronic 
ghost images, but these are always offset by 128 pixels from brighter 
sources, which is not the case here.} 
Moreover, the object is detected in two independent, ground--based 
near--IR data sets.   It is faintly visible ($S/N \approx 4$) in the 
$K_s$ (2.16$\mu$m) HDF images which we obtained with IRIM on the KPNO 4m 
telescope in 1996 April--May (cf.\ Dickinson 1998).  
\jdrop\ is the brightest of the five $K$--band selected, optically 
invisible candidates identified by Lanzetta \etal (1998) from the 
KPNO IRIM HDF images.  Their other four candidates have no counterparts 
in the NICMOS \J110 or \H160 data, implying either that they are 
exceptionally red at $\lambda \simgt 1.8\mu$m, or that they are not real.

To verify the $K_s$ detection and to improve the quality of the photometry, 
we obtained new $K_s$ images using NIRC (Matthews \& Soifer 1994) on the 
Keck~I telescope on UT 1999 April 5.  108 dithered 60~second exposures 
were taken through light, intermittent cirrus and processed using 
the DIMSUM\footnote{Deep Infrared Mosaicing Software, a package of 
IRAF scripts by Eisenhardt, Dickinson, Stanford and Ward, available
at ftp://iraf.noao.edu/contrib/dimsumV2.} reduction software.  The
individual frames were scaled and weighted before combination using 
measured counts of a moderately bright, nearby star centered
in the NIRC field.  The combined image has a PSF FWHM~$= 0\secspt5$.
The photometric zeropoint was bootstrapped from the IRIM $K_s$ data using 
large aperture measurements of two stars, which gave excellent internal 
agreement.  \jdrop\ is readily visible as a compact source in the 
NIRC image (Figure 1).

We summarize photometry of the object in Table~1.  In order to set 
limits in the HST bands \U300 through \J110, we measured fluxes in 
a $0\secspt7$ diameter circular aperture at the \H160 centroid position.  
This aperture maximizes $S/N$ for the object in the \H160 image, and is very 
close to that required to maximize $S/N$ for point sources.   We also have 
verified that the optical measurements in Table~1 are consistent with limits 
derived from the original, unconvolved WFPC2 images.  A 21\% aperture 
correction, derived from the \H160 image, has been applied 
to the measurements, errors and limits in all HST bands to adjust the fluxes 
to ``total'' values.  This correction assumes that the object has similar
morphology at all wavelengths, which cannot be verified at present.
We followed a similar procedure for the NIRC $K_s$ image, using photometry 
measured within a $1\secspt2$ diameter aperture, corrected to total flux 
using curve of growth measurements of the reference star.  We also 
measured the object in the IRIM $K_s$ image using a method based 
on that of Fern\'andez--Soto, Lanzetta \& Yahil (1999).  The NICMOS 
image is used as a template, convolved to match the IRIM PSF, and then 
fit to the ground--based data.   The NIRC and IRIM measurements 
($K_{AB} = 23.9$ and 23.8, respectively) agree within their 
errors, and with the value measured by Lanzetta \etal (1998) 
($K_{AB} = 23.7$) from the same IRIM data.  The $1\sigma$ and 
$2\sigma$ color limits for $\J110 - \H160$ are $>$2.8 and $>$2.3, 
respectively, while $\H160 - K_s = 1.23_{-0.19}^{+0.23}$.

\jdrop\ has $S/N < 2$ in all bands other than \H160 and $K_s$, regardless 
of the aperture size used.  However, we do measure formally positive 
counts above the background at \B450, \V606, \I814 and \J110.
Although measurements at low significance levels are prone to systematic 
uncertainties in background subtraction, etc., there is nevertheless the 
possibility that the object does have non--zero optical flux.  To explore this 
further, we have used an additional 63000s WFPC2 \I814 image of the HDF obtained 
by Gilliland, Nugent, \& Phillips (1999) in a search for high redshift 
supernovae.  R.~Gilliland kindly provided his sum of all available HDF--N 
$I_{814}$ data in the WF3 region, with a total exposure time of 186600s.  
We registered this image with the data in other bandpasses.  Photometry 
on the unconvolved, ``grand sum'' \I814 image in an $0\secspt7$ aperture 
centered at the nominal position of \jd1\ measures a flux of 
$5.7 \pm 2.6$~nJy (with no aperture correction), or a formal $S/N = 2.2$.  

To assess the significance of this optical measurement, we carried out 
a simple fluctuation analysis on the WFPC2 data.  The images were normalized 
to constant variance over the field of view by dividing by smoothed noise 
maps which are generated as part of the data reduction process.  
We used SExtractor to generate masks which exclude all readily detectable 
sources in each image, including generous buffer regions around each.  
We fit smooth background maps to the ``blank sky'' regions and subtracted 
them from the images.   The data were filtered by a Gaussian with 
FWHM~$= 0\secspt14$ (matching the WFPC2 PSF), and the distribution of pixel 
values over an 855~arcsec$^2$ region was compared to that near the position 
of \jdrop.  We did this for each WFPC2 image, for variance weighted sums of 
the optical bandpasses, and also for $\chi^2$ combinations 
constructed following the procedure of Szalay, Connolly, \& Szokoly (1999).  
The results from the $\chi^2$ images were statistically similar to those 
from the weighted sums, so we describe only the latter here.  The filtered 
pixel values have a nearly Gaussian distribution (whose width in a given
image we will characterize as $\sigma$ here), with a positive tail due to
faint sources.  At the nominal \jd1\ peak position, the filtered 
\I814 image has a value which exceeds the local background by $1.6\sigma$,
and in the weighted $\V606 + \I814$ image by $1.9\sigma$.  There is a 
stronger local peak in the filtered \I814 and \V606+\I814 images located 
$0\secspt14$ from the nominal \jd1\ position, which exceeds the local 
background by 3.1$\sigma$ in \I814 and by 3.3$\sigma$ in $\V606 + \I814$.
Including the \B450 data in the weighted sums or $\chi^2$ images does 
not strengthen the peak.  Its formal significance depends on how one 
treats the non--Gaussian tail of positive pixel values due to faint 
sources.  Excluding buffered regions around all objects detected in the 
infrared catalogs, we find that there is a 5.1\% chance of a pixel in the 
filtered \V606+\I814 images exceeding this ``3.3$\sigma$'' threshold within 
$0\secspt14$ of a location specified {\it a priori}.  The infrared catalogs, 
however, miss faint blue galaxies which are readily visible in the optical 
data, and which contribute to this positive tail.  Excluding all objects 
detected in the infrared {\it or} optical SExtractor catalogs, the chance 
probability drops to 1.8\%.  The exact probabilities depend on the detection 
threshold used for the catalogs.  We conclude that the optical ``detection''
corresponds to a Gaussian significance of 1.6 to 2.2$\sigma$.

\jdrop\ is not detected in the VLA radio maps of Richards \etal (1998)
and Richards (1999), with a $3\sigma$ limit of 4.8$\mu$Jy at 8.5~GHz
and 22.5$\mu$Jy at 1.4~GHz.  It is not reported as a source in the 
ISO HDF catalogs of Goldschmidt \etal (1997), Aussel \etal (1999) or 
D\'esert \etal (1999), with approximate flux limits of 50$\mu$Jy and 
25$\mu$Jy at 7 and 15$\mu$m,\footnote{These ISO flux limits are estimated 
based on measurements reported by Aussel \etal (1999) for two other
``supplemental'' 15$\mu$m sources in the same general region of the HDF.} 
nor as an 850$\mu$m source in the SCUBA observations of Hughes \etal (1998), 
with a limit of 2.0 mJy (4.4$\sigma$).  Interestingly, the next reddest
HDF object, J123651.74+621221.4 (with $\J110 - \H160 = 1.6$, $\H160 = 24.3$;  
see Figure 2), corresponds to a VLA/MERLIN radio source, and may also have 
a 15$\mu$m counterpart in the Aussel \etal (1999) ISO source list, as well 
as a possible $3\sigma$ detection in the 1.3mm IRAM map of Downes \etal (1999).

\section{Spectroscopy}

We observed \jdrop\ using the Cryogenic Spectrograph (CRSP; Joyce 1995) 
at the KPNO 4m on UT 1999 May 2, 3, 6 and 7.   We acquired the target using 
a blind offset from a bright star located in the HDF flanking fields.
The observations were taken through a 1\secspt0 wide slit oriented at 
PA~=~119\degspt4 to cover both the target object and a nearby star
(12\secspt8 away) which served as a pointing reference.  The target 
was dithered along the slit in an ABBA pattern using 100~second exposure 
times, and the count rate from the reference star was monitored to ensure 
that pointing remained stable.  The target was periodically reacquired and 
placed at new positions along the slit.  Seeing throughout all 
observations averaged 1\secspt0.  On 2 and 6 May we used grating 4
(200 lines/mm, blazed at 3$\mu$m) at 1st order in the $K$--band
($\lambda\lambda 1.90$--$2.50\mu$m) giving an effective spectral 
resolution $R = \lambda/\Delta\lambda \approx 240$ at $\lambda 2.2\mu$m 
as measured from night sky lines.  On 3 and 6 May we also observed with 
grating 4 at 2nd order in the $H$--band ($\lambda\lambda 1.49$--$1.80\mu$m), 
with $R \approx 380$ at $\lambda 1.65\mu$m.  The $H$--band data on 3~May 
were taken through occasional cirrus;  frames with poor transparency 
(judged from the reference star) or bad sky subtraction were discarded 
from the analysis.  The total exposure times retained for the final $H$ 
and $K$--band grating 4 spectrograms were 13400s and 17600s, respectively.
We reduced the data using standard procedures.  The data were corrected 
for array non--linearity, and flat--fielded using dome flats.  The sky 
was subtracted (to first order) using the ABBA differences.  The wavelength 
scale and geometric distortion were calibrated using OH night sky lines, 
and the two dimensional spectral images were rectified.  Residual sky 
features were then fit and subtracted.  The images were positionally 
registered using the reference star, and combined using a bad pixel
mask and a percentile--clipping scheme to reject outlying pixels.  
Spectra were extracted through a 3 pixel (1\secspt05) wide window 
at the nominal position of the object relative to the reference star, 
along with a noise measurement extracted from the 2--dimensional variance
image created when the images were combined.  Flux calibration was
based on observations of the Elias \etal (1982) standard HD 105601,
with absolute flux normalization based on the $H$ and $K_s$ magnitudes 
(measured from the IRIM HDF data) of the reference star on the slit.

The low--resolution $H$--band spectrogram (Figure 3) shows 
a possible emission line at $\lambda 1.643\mu$m, with 
flux~$\approx 2\times10^{-16}$~erg~s$^{-1}$~cm$^{-2}$ and integrated
$S/N \approx 4.3$.  Although the $H$--band sky is nearly covered with 
OH emission bands at this low dispersion, the telluric lines near this 
wavelength are relatively weak.   We reobserved the object at higher 
spectral resolution ($R \approx 670$ at $\lambda 1.65\mu$m) on UT 1999 
May 7 under good atmospheric conditions, using grating 1 (300 lines/mm, 
blazed at 4$\mu$m) in 2nd order, covering $\lambda\lambda$1.51--1.69$\mu$m.
The putative emission line does not reproduce in the grating 1 spectrogram.
The $3\sigma$ upper limit for an unresolved emission line at 1.643$\mu$m 
is $\approx 6 \times 10^{-17}$~erg~s$^{-1}$~cm$^{-2}$.

Careful, frame--by--frame inspection of the grating 4 data does not 
reveal any obvious artifacts that could have produced the emission 
feature.  It is present when the data are combined as a median 
(rather than percentile clipped averages), demonstrating that the 
feature results from systematically high data values and not from 
intermittent artifacts.   We have tried to test its reality by 
subdividing the grating 4 exposures into independently averaged, 
randomly selected half data sets.  However the $S/N$ in each 
half--set is low;  the apparent line sometimes appears in both, 
sometimes not.  Its significance is hard to evaluate because 
noise in the data is correlated (e.g., sky subtraction residuals) 
with greatly varying amplitude vs.\ wavelength.  As a test, we 
normalized the coadded, two dimensional spectral image by the variance 
map to equalize the pixel--to--pixel RMS throughout, and extracted 
26 spatially independent, 3 pixel wide spectra from regions of the 
slit unaffected by the reference star.  Each extraction was subtracted
to zero mean, and then convolved by a Gaussian with the instrumental 
resolution.  We measured the peak value of the nominal emission line 
from \jd1, and searched for features with equal or greater amplitude 
(positive or negative) at any wavelength in the other 25 ``test'' 
extractions.  No comparable positive feature was found, and only 
one negative feature, suggesting a probability $\sim 2\%$ 
(1 out of $2 \times 25$) that the \jdrop\ ``line'' would arise 
by chance.   

Our spectroscopic results are therefore ambiguous.  The line in 
the grating 4 spectrogram is resilient but not ironclad.  The fact that 
it does not reproduce in the grating 1 observation suggests either that 
it is not real, or that it is well resolved at the higher dispersion 
(thus reducing the detection sensitivity), or that the object was not 
well placed in the spectrograph slit on May 7.

The $K$--band spectrogram shows no significant emission features;  
the $3\sigma$ flux limit for an unresolved line ranges from 
1 to $3\times10^{-16}$~erg~s$^{-1}$~cm$^{-2}$
over the range $\lambda\lambda1.95$--2.4$\mu$m, and rises steeply
at longer wavelengths due to the increasing thermal background.  
We would not have expected to (and did not) detect the object's 
continuum at either $K$ or $H$.  

\section{Angular Extent}

The NICMOS F160W image of \jdrop\ is very compact, but there is 
evidence that it may be resolved.   Examining ten well exposed, 
spectroscopically confirmed stars in our images, we measure the PSF FWHM 
to be $0\secspt217 \pm 0\secspt016$ (0\secspt19 to 0\secspt24 maximum range), 
while \jd1\ has FWHM~$= 0\secspt28$.  Figure 4a compares the surface
brightness profiles of the object and three faint, well isolated stars 
which have been registered and scaled using a non--negative 
cross--correlation procedure.  The profile of \jdrop\ appears to 
be slightly more extended.  Subtracting the scaled PSF stars from 
\jd1\ leaves positive annular residuals which are 
not present when one star is scaled and subtracted from another.   
Additionally, SExtractor computes a stellarity index using a neural 
network classifier which outputs values from 0 (extended) to 1 (stellar).
The known HDF stars have stellarity $\geq 0.84$, while \jd1\ has 
stellarity $= 0.08$ (Figure 4c).  However, the object is near the faint 
limit where the classifier appears to be reliable.  The NICMOS PSF
may depend on source color, and \jdrop\ is extremely red while the 
stars are among the bluest objects in $\J110 - \H160$.  We have 
examined this using Tiny Tim models (Krist \& Hook 1997) computed
for a very wide range of source spectra and find it is too weak
to account for the differences we measure.  Proper analysis of the 
PSF from dithered images with NICMOS Camera 3 requires a realistic 
treatment of the known (and large) detector intrapixel sensitivity 
variations (cf.\ Lauer 1999).   Therefore while our measurements 
suggest that the object is spatially extended, we cannot be completely 
confident about this given the present data.  In the future, a robust 
angular extent measurement might be accomplished using NICMOS Camera 2,
which critically samples the PSF at 1.6$\mu$m, when the instrument 
is resuscitated with the NICMOS Cooling System in HST Cycle~10.

Figure~5 compares the colors of \jdrop\ to those of a variety of cool
or reddened stars.   We have converted our NICMOS AB magnitudes 
to a conventional Vega scale for comparison to published stellar data, 
including an approximate color term to correct to standard $J$ and $H$ 
bandpasses.  In this system, \jd1\ has $J-H \simgt 2.5$ (2$\sigma$) 
and $H-K \approx 1.6$.   This is much redder than ordinary cool stars 
and known substellar objects, which can have extremely red 
optical--to--infrared colors but are generally rather blue in $J-H$ 
and $H-K$ where their spectra are dominated by strong molecular 
absorption.   The reddest L dwarf stars reach $J-H=1.45$, but the
known ``methane dwarfs'' have $J-H \approx H-K \simlt 0$.  
Atmosphere models (Burrows \etal 1997) predict still bluer colors 
for cooler brown dwarfs and giant extra--solar planets.   
The only known stars as red as \jd1\ are those undergoing 
mass loss with thick circumstellar dust shells, such as extreme 
carbon stars or Mira variables.  \jdrop\ is redder than ordinary 
galactic or Magellanic AGB stars, but the most heavily reddened 
objects like IRC~+10216 and some extreme Miras can equal or exceed 
its colors.\footnote{IRC~+10216 itself, scaled to the \H160 or $K_s$ 
magnitude of \jdrop, should have been detected in the ISO images 
of the HDF at 7 and 15$\mu$m.} 
It would be remarkable to encounter such an unusual star in a tiny, 
high galactic latitude field such as the HDF.  Moreover, at $K$
it is 11.9 magnitudes fainter than the reddest LMC carbon stars,
implying (by analogy) a distance $\sim 12$~Mpc.
A blackbody with $T \approx 1050$K also matches the $\H160 - K_s$ color 
and $2\sigma$ $\J110 - \H160$ limit for \jdrop\ quite well, but
no known star has such a spectrum.   Recent atmosphere
models for cool white dwarfs (e.g., Hansen 1998;  Saumon \& Jacobson 1999)
suggest that they should have very blue colors due to $H_2$ opacity.   

\section{Discussion}

The salient photometric features \jdrop\ are firm detections at \H160 
and $K_s$, and very faint upper limits (possibly with some $S/N \simlt 2$
detections) at all shorter wavelengths.   The bandpass--averaged 
$f_\nu$ flux density declines by a factor of 3.1 from 2.2$\mu$m to 
1.6$\mu$m, and then again from 1.6$\mu$m to 1.1$\mu$m by a factor 
$34^{+\infty}_{-21}$ (1$\sigma$ errors), with a 97.7\% (1--sided) 
confidence limit $> 8.3$, implying curvature or a break in the 
spectral energy distribution.  This is the only object in the 
NICMOS HDF--N at $\H160 < 26$ which is undetected (with $S/N < 2$) 
in any of the optical WFPC2 bandpasses.   It is by far the reddest 
HDF--N object in $\J110 - \H160$, but also one of the reddest 
in $\H160 - K_s$.  Only one (slightly) brighter HDF object, 
a $z = 3.2$ ``$U$--dropout'' galaxy, has a redder $\H160 - K_s$ 
color (but equal within the $1\sigma$ measurement uncertainties).  
That galaxy's color probably results from a strong Balmer and/or 
4000\AA\ break, or possibly from strong [OIII]+H$\beta$ line 
emission in the $K_s$ band.

There are three common reasons why a galaxy's spectral energy 
distribution (SED) may appear to be very red:  extinction, age, 
or the presence of a strong spectral break, such as that caused 
by the Lyman limit or Lyman~$\alpha$ forest.   Here we consider
each of these in turn, making comparisons wherever possible to 
photometry of actual galaxies from the HDF and elsewhere in order 
to avoid overreliance on models.  One way of comparing colors in 
different bandpass combinations is to parameterize the photometry 
by a spectral index $\alpha$, i.e. $f_{\nu} \propto \nu^{-\alpha}$.
For two bandpasses with effective wavelengths $\lambda_1$ and $\lambda_2$ 
and magnitude measurements $m_1$ and $m_2$, we may define 
$\alpha_{(m_1-m_2)}  = 0.4(m_1 - m_2) / \log(\lambda_2 / \lambda_1)$.
The nominal \jdrop\ $\J110 - \H160$ color corresponds to
$\alpha_{(J-H)} = 10.2$, and the $1\sigma$ and $2\sigma$ 
limits to $\alpha_{(J-H)} > 7.5$ and $> 6.1$, respectively.
Considering pairs of adjacent HDF bandpasses \U300 through \J110, 
we find a few other galaxies with comparably steep spectra at shorter 
wavelengths.  Among these, the objects with spectroscopic identifications 
are invariably either Lyman break ``dropouts,'' or red, early--type 
galaxies, and the as--yet unidentified objects all appear to be 
consistent with being members of one of these two classes.

Dust may, in principle, redden a spectrum almost arbitrarily.  
Because extinction laws generally steepen in the UV, a red 
and possibly strongly curved SED may result when the UV portion of 
a distant, dusty galaxy redshifts into optical or near--IR bandpasses.
One well--known and fairly extreme example is the object discovered 
by Hu \& Ridgway (1994) and colloquially known as HR10, with $z=1.44$ 
(Graham \& Dey 1996).  Sub--mm measurements by Cimatti \etal (1998) and 
Dey \etal (1999) have shown that it is probably a dust--enshrouded,
star forming galaxy or AGN.   The SED of HR10 is steepest between 
$I$ and $J$ (Dey \etal 1999), with $\alpha_{(I-J)} = 6.1$.  An object 
like HR10 shifted to $z \approx 2.3$ would have colors roughly 
consistent with our measured 2$\sigma$ limits on \jdrop\ (see Figure~6).
A somewhat redder SED might be needed to fully match the photometry 
of \jd1, but this could presumably be accomplished by adding still 
more dust.  HR10 is compact in Keck $K$--band images, but diffuse 
($\approx 1\secspt2$ in size) and bimodal in WFPC2 images which 
sample the rest--frame UV light (Dey \etal 1999).  Similar changes 
in morphology with wavelength could make it difficult to detect
a fainter, more distant analog to HR10 in the WFPC2 HDF.  
Adopting (here and below) a cosmology with 
$\Omega_M = 0.3$, $\Omega_\Lambda = 0.7$, and 
$H_0 = 70$~km~s$^{-1}$~Mpc$^{-1}$, 
at $z \approx 2.3$ \jdrop\ would have a rest--frame $B$--band 
luminosity $\approx 0.12\times$ that of HR10.  Scaling the far--IR 
emission from HR10 accordingly, the predicted 850$\mu$m flux would 
be $\sim$0.5~mJy, or a factor of four below the current SCUBA limits
for the HDF.  Objects as red as HR10 are relatively rare, and it 
may seem surprising to find an even redder example within the 
5~arcmin$^2$ of the HDF, but little is known about their numbers 
at these faint magnitudes.  The fact that there are only a handful
of objects with such extreme IR colors in the HDF suggests that they
are not a common population by number relative to ordinary, relatively
unreddened galaxies, but their unobscured luminosities and star 
formation rates might be quite large and important in the scheme 
of galaxy formation.  Some of the faint sub--mm sources detected 
in recent surveys are red objects like HR10 (e.g., Smail \etal 1999), 
although it is worth noting that none of the candidate counterparts 
to sub--mm sources in the HDF has particularly unusual colors 
(Dickinson \etal 1999).

The integrated spectrum of an old stellar population is steepest 
in the near ultraviolet, from the familiar 4000\AA\ break through
spectral breaks at 2900\AA, and 2640\AA\ (Morton \etal 1977).  
At $2.5 \simlt z \simlt 4.5$ these breaks would redshift beyond
the WFPC2 bandpasses into the near--IR, maximizing the $\J110 - \H160$ 
color.  The resulting $k$--correction could make old, high redshift 
ellipticals nearly invisible in the optical HDF (e.g., Maoz 1997).   
For example, the observed--frame colors of the giant elliptical galaxy 
HDF 4-752.1 at $z = 1.013$ correspond to $\alpha_{(B-V)} \approx 10.2$ 
and $\alpha_{(V-I)} \approx 6.3$.  Shifted to $z \simgt 3$, its SED 
would roughly match that of \jdrop\ (Figure~6).  However, this would 
require that a $z \approx 3$ elliptical have a spectrum nearly as red as 
that of 4-752.1 at $z = 1$, despite the universe being considerably younger.  
This is difficult to explain without invoking an unusual IMF, an unpopular 
cosmology, or extinction.  Figure~7 shows the infrared colors of HDF 
galaxies with known redshifts, along with models computed using the 
population synthesis code of Bruzual \& Charlot (1993, 1996).  The 
reddest model traces a solar metallicity population formed with 
a Salpeter IMF in a short burst at $z=15$ and aging passively 
thereafter.  The $\H160 - K_s$ color of \jdrop\ is reasonably matched 
for $3 < z < 6$, but the $\J110 - \H160$ color of the model does not 
quite reach the formal $2\sigma$ color limit for \jd1\ at any redshift.
At $z = 3.5$ in our adopted cosmology, \jd1\ would have an absolute 
$B$ magnitude $M_B \approx -21.6$.
If it were an ``old'' elliptical galaxy at $z \simgt 3$, 
this would strongly suggest that at least some galaxies formed the 
bulk of their stars at extremely large redshifts.
If so, however, then it is puzzling that there are no other 
HDF galaxies nearly as red, as might be expected if there were 
a continuum of such objects extending out to $z \approx 3$.  Red, 
early--type galaxies have been identified spectroscopically out 
to $z = 1.55$ (e.g., Dunlop \etal 1996;  Spinrad \etal 1997;
Dickinson 1997;  Dey 1998) and photometrically in the HDF 
and elsewhere out to $z \approx 2$ (cf.\ Stiavelli \etal 1999, 
Ben\'{\i}tez \etal 1999), but \jdrop\ would stand alone as a unique 
example at significantly larger redshift.  

The Lyman limit can effectively truncate the spectrum of high redshift 
objects, making them appear arbitrarily red in certain color combinations.  
The $z > 5$ ``V--dropout'' galaxies HDF 4-473 and 3-951 (Weymann \etal 1998, 
Spinrad \etal 1998) and the $z=4.022$ ``B--dropout'' HDF 3--512 (Dickinson 1998) 
all have $\alpha > 7.3$ for colors that span the Lyman limit, and 
several $z \sim 3$ ``U--dropouts'' have $\alpha_{(U-B)} > 6$.  
At $z \simgt 3$, an additional break is introduced by the Lyman~$\alpha$ 
forest, which is increasingly thick at higher redshifts and affects 
broad band colors to a correspondingly greater degree 
(cf.\ Lowenthal \etal 1997, Dickinson 1998).  We do not know 
the opacity of the Lyman~$\alpha$ forest at $z > 6$, but may reasonably
assume that it continues to increase since we believe that the
universe was dominated by neutral hydrogen beyond some reionization redshift.

If we interpret \jdrop\ as a Lyman break object, then its $\J110-\H160$ 
color is matched for $z \simgt 10$, but the red $\H160-K_s$ color suggests 
that the Lyman~$\alpha$ forest has eaten away roughly half of the $\H160$ 
flux, implying a redshift $z \approx 12.5$ (see Figure~7).  At that redshift, 
the $K$--band corresponds to $\lambda_0 = 1600$\AA\ in the emitted frame,
essentially the same wavelength where Steidel \etal select galaxies in 
their survey at $z \approx 3$.   We may therefore make a direct comparison 
to the observed $z \approx 3$ luminosity function (Dickinson 1998, 
Steidel \etal 1999).  For an $\Omega_M = 0.3$, $\Omega_\Lambda = 0.7$ 
cosmology, \jdrop\ would be $\approx 3\times$ more luminous than the 
brightest Lyman break galaxy from our 0.3~deg$^2$ ground--based survey.
This may seem improbable, given the small solid angle of the HDF.  
However, the total co--moving HDF volume out to $z = 12.5$ is quite 
large, particularly for an open or $\Lambda$--dominated cosmology.  
For $\Omega_M = 0.3$, $\Omega_\Lambda = 0.7$, it is equivalent to the 
effective volume of the $U_nG{\cal R}$ Lyman break selection function 
(see Steidel \etal 1999) over a 100~arcmin$^2$ field.  Because the volume 
is heavily weighted toward the highest redshifts, then it is not unlikely 
that the rarest, most luminous objects would also be the most distant, 
provided that they exist at all beyond $z > 6$.  In our ground--based 
survey we often find $z \approx 3$ QSOs that are brighter than the most 
luminous Lyman break galaxies, and the nearly unresolved morphology 
of \jd1\ might plausibly indicate that it is some sort of AGN.  
Populations of distant QSOs have been postulated as a means of reionizing 
the universe at high redshift, although existing models for the formation 
of high redshift AGN (Haiman \& Loeb 1998;  Haehnelt, Natarajan, \& Rees 1998)
would predict that no objects as bright as \jd1\ should be found within 
the HDF volume at $z > 10$.

If \jdrop\ were a galaxy forming stars at $z = 12.5$ with a Salpeter IMF 
and without dust, its UV luminosity would correspond to a star formation 
rate~$\approx 180h_{70}^{-2} M_\odot/{\rm yr}$ 
(70 or 400$h_{70}^{-2} M_\odot/{\rm yr}$ for Einstein--de~Sitter 
or $\Omega_M = 0.2$ open universes).  This is the sort of rate required 
by monolithic collapse models which would form a $10^{11} M_\odot$ galaxy
within a short ($\simlt 10^9$~yr) time scale, and is also comparable to 
the {\it obscured} star formation rates which have been claimed for high
redshift sub--mm sources.  However, if \jdrop\ resembles a classical, 
unobscured protogalaxy, then such objects are evidently quite rare.   
We will present a more complete discussion of color--selected high 
redshift galaxy candidates from our NICMOS survey in a future paper, 
but there are no $U$, $B$, $V$ or $I$ ``dropout'' candidates in the HDF 
with luminosities comparable to that which \jd1\ would have if it were 
at $z = 12.5$.   One such object in the redshift range $2 < z < 12.5$
implies a space density $\sim 10^{-5}h_{70}^{3}$~Mpc$^{-3}$ for our 
assumed cosmology, and $4.3\times$ larger for an Einstein--de~Sitter 
universe.  This is a few hundred times smaller than the present--day 
space density of $\sim L^\ast$ galaxies ($\phi^\ast = 4$ to 
$6h_{70}^{3} \times 10^{-3}$~Mpc$^{-3}$ from the $K$--band luminosity 
functions of Mobasher \etal 1993, Gardner \etal 1997, Szokoly \etal 1998,
and Loveday 1999).  If rapid, monolithic galaxy formation took place 
anywhere in that redshift range, then either it was quite uncommon,
or it was obscured by dust.  Lanzetta \etal (1998) reached very similar 
conclusions based on their search for optically invisible objects in 
the KPNO infrared HDF data.  If perhaps only the rarest, most massive 
galaxies, e.g., brightest cluster ellipticals (BCEs), formed their
stars rapidly at the highest redshifts, then \jdrop\ could be one 
example.  The present--day space density of galaxy clusters with 
Abell richness class $\geq 1$ or X--ray temperature $kT \geq 2.5$~keV
is $\approx 3 \times 10^{-6}h_{70}^{3}$~Mpc$^{-3}$ 
(cf.\ Eke, Cole, \& Frenk 1996), implying that we might
expect $\sim 1/3$ ``proto--BCEs'' per HDF volume.

From Figures 7 and 1 it is evident that a star--forming galaxy at 
$z \simgt 9.5$ should have $\J110 - \H160 \simgt 1.5$ due to the 
Lyman limit, and also that \jdrop\ is the only such object in the 
HDF--N with $\H160 < 26$, the magnitude down to which we could measure 
such a color or limit with $> 2\sigma$ significance.\footnote{The 
radio source counterpart J123651.7+621221.4 is well
detected in the \B450, \V606 and \I814 optical passbands, and 
thus is not at $z \gg 3$.}  There are, therefore, no other 
{\it detected} candidates for galaxies at $z > 9.5$ in the HDF, 
although cosmological surface brightness dimming could significantly 
affect sensitivity to extended protogalaxies at such redshifts, 
even in deep images such as these.   At $z = 9.5$, the $\H160 = 26$ 
limit corresponds to an unobscured star formation rate 
$\approx 20h_{70}^{-2} M_\odot$~yr$^{-1}$ for our adopted cosmology.

The possible 1.643$\mu$m emission line seen in the CRSP spectra could 
be consistent with any of these interpretations.   For a dusty galaxy 
or AGN, it might be [OIII]$\lambda$5007\AA\ at $z=2.28$.  H$\alpha$ 
at $z \approx 1.5$ is also possible, but it is much easier to match
the colors at $z \simgt 2$ where the $\J110$ band would sample the 
rest--frame UV.   Alternatively the line could be [OII]$\lambda$3727\AA\
at $z=3.40$, a redshift where an old stellar population nearly matches 
the colors of \jdrop.  An emission line, however, might suggest active 
star formation inconsistent with a maximally old elliptical galaxy, 
thus requiring either the presence of an AGN, dust, or both.  A reddened,
star forming object at this redshift is also possible.  Finally, the line 
could be Lyman~$\alpha$ at $z=12.51$, where the $JHK$ colors are well 
matched by the Lyman break hypothesis.   However, given the fact that 
the line did not reproduce in our reobservation (see \S3 above), 
we cannot be confident that it was real and do not further consider 
the spectroscopic possibilities here.

An alternative explanation for the peculiar colors would be the 
presence of very strong line emission in one or more infrared
bands.  Line fluxes $\sim1.5$ and $2 \times 10^{-16}\ergpscm2$
in the $\H160$ and $K_s$ bands, respectively, could mimic the broad
band fluxes measured for \jdrop.  The possible 1.643$\mu$m emission 
line, if real, could therefore account for the entire signal 
detected in the NICMOS \H160 image.  Detection at both $\H160$ and 
$K_s$ makes the ``pure emission line'' hypothesis seem less plausible,
however, requiring either strong lines in both bands or an extremely 
red continuum.  In the latter case, we are forced back to our previous 
speculations.

\section{Conclusion}

Without further data, we cannot unambiguously distinguish between
these explanations for the nature of \jdrop.  Each, however, is 
quite remarkable in its own right.  Perhaps the least ``exotic'' 
extragalactic hypothesis is that \jd1\ is a dusty, HR10--like galaxy 
at $z \simgt 2$.  Such objects, with sub--mm fluxes just below the
current SCUBA detection limits, might be sufficiently common
to make up the bulk of the far--infrared background (Barger, 
Cowie \& Sanders 1999).  The possibility that \jdrop\ is an ``old'' 
elliptical galaxy at $z \simgt 3$ is perhaps more remarkable, 
in that it would strongly suggest that at least some galaxies formed 
the bulk of their stars at very large redshifts.\footnote{When
first forming stars, the $z > 10$ progenitor to an old, $z \approx 3.5$ 
elliptical might well resemble \jdrop.}  

Perhaps the most spectacular interpretation of this object, but also
the one requiring the most rigorous proof, would be that it lies at 
$z > 10$, an extremely distant analog of the Lyman break galaxy 
population.  If this were the case, then its high luminosity and the 
relatively small volume of the HDF would suggest that we were either 
extremely lucky, or that such objects were more common than would be 
expected based on an extrapolation of the $z \approx 3$ population 
to higher redshifts.  If the luminosity results from star formation, 
then \jdrop\ would resemble the classical picture of a protogalaxy, 
forming stars at $\simgt 100 M_\odot/{\rm yr}$.  However, the 
implied space density of {\it unobscured} objects with such star formation 
rates is far smaller than that of $L^\ast$ galaxies today.  If galaxies 
formed monolithically at high redshift, then this was either a rare 
occurrence, or the process was largely enshrouded by dust, as has 
been suggested by the recent detection of distant sub--mm sources.  
If, instead, \jd1\ were a QSO, it could be part of the population of 
objects responsible for re--ionizing the universe.  From our NICMOS
HDF data, we should be able to identify unreddened, compact objects 
with star formation rates $\sim 20h_{70}^{-2}M_\odot/{\rm yr}$ at 
$z \approx 9.5$, with the limit rising to 
$\sim 100h_{70}^{-2}M_\odot/{\rm yr}$ at $z=13$ as the 
Lyman~$\alpha$ forest suppresses the \H160 flux.  
No candidates other than \jdrop\ are seen.

Distinguishing between the possible explanations for this object will 
require new observations, but none will be easy.  Spectroscopy will 
be challenging even with the new generation of IR instruments on 8 to 
10m telescopes, and there is no guarantee that the object has detectable 
emission lines.  Although our understanding of the transparency of the 
IGM at $z > 5$ is based solely on extrapolation, a robust optical 
detection would probably exclude the Lyman break hypothesis.  
Occasionally, QSO sightlines at lower redshifts are free of optically 
thick H~I absorption (e.g., Reimers \etal 1992), but it seems unlikely 
that emitted--frame Lyman continuum from an ordinary galaxy or QSO 
could propagate through the universe from $z \approx 12$ without being 
absorbed.   The optical measurement in \S2 is suggestive but not
highly significant;  a deeper optical image, e.g., with STIS or the 
forthcoming HST Advanced Camera for Surveys, would be a valuable 
(albeit expensive) observation.  Photometry at $\lambda \simgt 3\mu$m 
with IRAC on SIRTF could distinguish between the $z > 10$ Lyman break 
and $z < 6$ red galaxy hypotheses, since the spectra of red galaxies 
should continue to rise toward longer wavelengths, with 
$f_\nu \simgt 2\mu$Jy, while a star forming galaxy at $z > 10$ 
should have a flatter, fainter SED (see Figure~6).  If near--IR 
spectroscopy fails to detect emission lines, then the best hopes for 
distinguishing between the old galaxy and reddened starburst/AGN models 
lie at longer wavelengths.  Given that the HDF already has the deepest 
(and possibly confusion limited) SCUBA observation, a true sub--mm 
detection may require a future generation of telescopes and instruments.  
A detection would certainly imply dust, but given the negative sub--mm 
$k$--correction it probably would not distinguish between ``low'' 
($z \sim 2$) and high ($z \sim 12$) redshifts without multifrequency 
measurements.  For a reddened starburst at $z \sim 2.3$, the dust 
emission should peak near $\sim 200\mu$m and might be detectable 
by the SIRTF MIPS instrument near its confusion limit.  SIRTF IRAC 
photometry from 3.6 to 8$\mu$m may prove useful, although disentangling 
reddening and age from broad band SEDs is notoriously difficult.

Regardless of its nature, the fact that such an unusual 
and extreme object was found in a 5~arcmin$^2$ field suggests that 
there are interesting surprises awaiting future, wide--field, deep 
infrared surveys.

\acknowledgements

We would like to thank the other members of our HDF--N/NICMOS GO team
who have contributed to many aspects of this program, and the STScI
staff who helped to ensure that the observations were carried out in
an optimal manner.  We also thank the support staff at KPNO and the 
W.M.~Keck Observatories for their help in carrying out the ground--based 
observations, especially Dick Joyce for verifying the CRSP plate 
scale.  We thank Ron Gilliland for providing his additional WFPC2 
images of the HDF, Eric Richards for digital VLA maps, and Sandy Leggett 
for digital tables of stellar photometry.  Arjun Dey and the editor, 
Greg Bothun, carefully read the manuscript and made helpful suggestions.
Jim Liebert, Michael Liu and William Bidelman provided guidance and 
references concerning colors of unusual galactic stars.   
Support for this work was provided by NASA through grant 
number GO-07817.01-96A from the Space Telescope Science Institute, 
which is operated by the Association of Universities for Research 
in Astronomy, Inc., under NASA contract NAS5-26555.

\clearpage

\begin{deluxetable}{ccccc}
\footnotesize
\tablewidth{0pt}
\tablecaption{Photometry of HDF--N~J123656.3+621322}
\tablehead{
\colhead{Instrument}
  & \colhead{Bandpass}
  & \colhead{Wavelength}
  & \colhead{Flux Density\tablenotemark{a}}
  & \colhead{AB Magnitude\tablenotemark{b}} \nl
}
\startdata
NIRC   & $K_s$  &  2.16$\mu$m  &  $995.0 \pm 182.5$ nJy  &  $23.91^{+0.22}_{-0.18}$ \nl
NICMOS & \H160  &  1.61$\mu$m  &  $318.1 \pm  13.8$ nJy  &  $25.14^{+0.05}_{-0.05}$ \nl
NICMOS & \J110  &  1.14$\mu$m  &  $  9.3 \pm  14.6$ nJy  &  $>27.44$  \nl
WFPC2  & \I814  &  0.80$\mu$m  &  $  6.1 \pm   4.1$ nJy  &  $>28.51$  \nl
WFPC2  & \V606  &  0.60$\mu$m  &  $  4.3 \pm   2.4$ nJy  &  $>29.00$  \nl
WFPC2  & \B450  &  0.46$\mu$m  &  $  1.5 \pm   3.6$ nJy  &  $>29.05$  \nl
WFPC2  & \U300  &  0.30$\mu$m  &  $ -2.5 \pm   7.5$ nJy  &  $>28.66$  \nl
\enddata
\label{jdropphot}
\tablenotetext{a}{Fluxes and uncertainties with aperture corrections to ``total''
values (see text).}
\tablenotetext{b}{When $S/N < 2$, magnitudes are quoted as $2\sigma$ limits.}
\end{deluxetable}

\clearpage

\begin{figure}
\plotfiddle{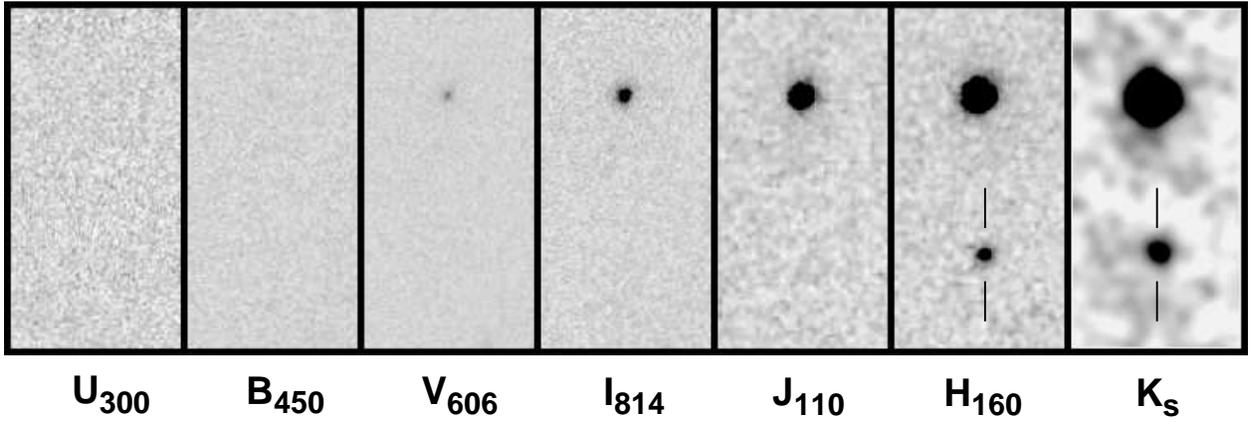}{3in}{-90}{80}{80}{-315}{400}
\caption{
%\figcaption[f01.ps]{
HST and Keck images of \jdrop\ at 0.3--2.16$\mu$m.  The field of
view of each panel is $4\arcsec \times 8\arcsec$, and north is
23\degspt8 counterclockwise from vertical.  \jdrop\ is identified 
by tick marks in the \H160 and $K_s$ panels, and is located at 
J2000 coordinates $\alpha = $~12:36:56.32, $\delta = $~62:13:21.7.
The NIRC $K_s$ image has been smoothed by a Gaussian with 
FWHM~$= 0\secspt38$.  A constant grey level corresponds
to constant $f_\nu$ surface brightness.  The elliptical galaxy at 
top (HDF 3-48.0) is very red in its own right;  its estimated 
photometric redshift is $z \approx 1.25$ 
(Budavari \etal 1999).
}
\end{figure}

\begin{figure}
\plotfiddle{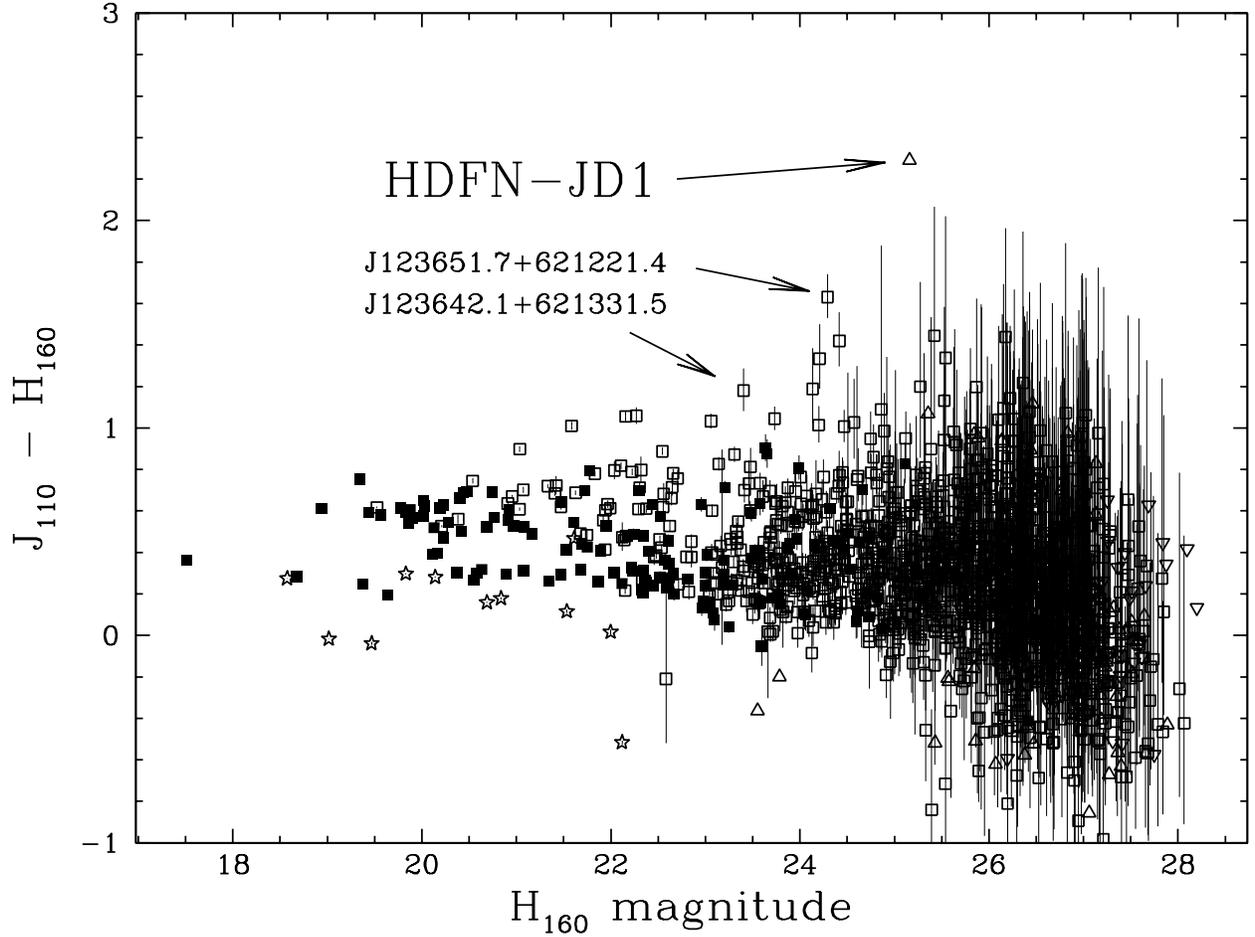}{6in}{-90}{65}{65}{-250}{400}
\caption{
%\figcaption[f02.ps]{
$\J110 - \H160$ versus \H160 color--magnitude diagram (on the
AB system) for the HDF--N NICMOS--selected catalog.    Filled and open 
squares indicate galaxies with and without spectroscopic redshifts, 
respectively, while known stars are marked by star symbols.  
Colors are plotted with $\pm 1\sigma$ error bars, and measurements 
with $S/N < 2$ are marked by triangles at the $2\sigma$
color limit.  \jdrop\ is labeled;  its $1\sigma$ color limit would be 
$\J110 - \H160 > 2.8$.  The next--reddest galaxy, J123651.7+621221.4, 
is a faint cm radio source.  Another VLA source, J123642.1+621331.5, 
is also unusually red (Richards \etal 1998, Waddington \etal 1999).
}
\end{figure}

\begin{figure}
\plotfiddle{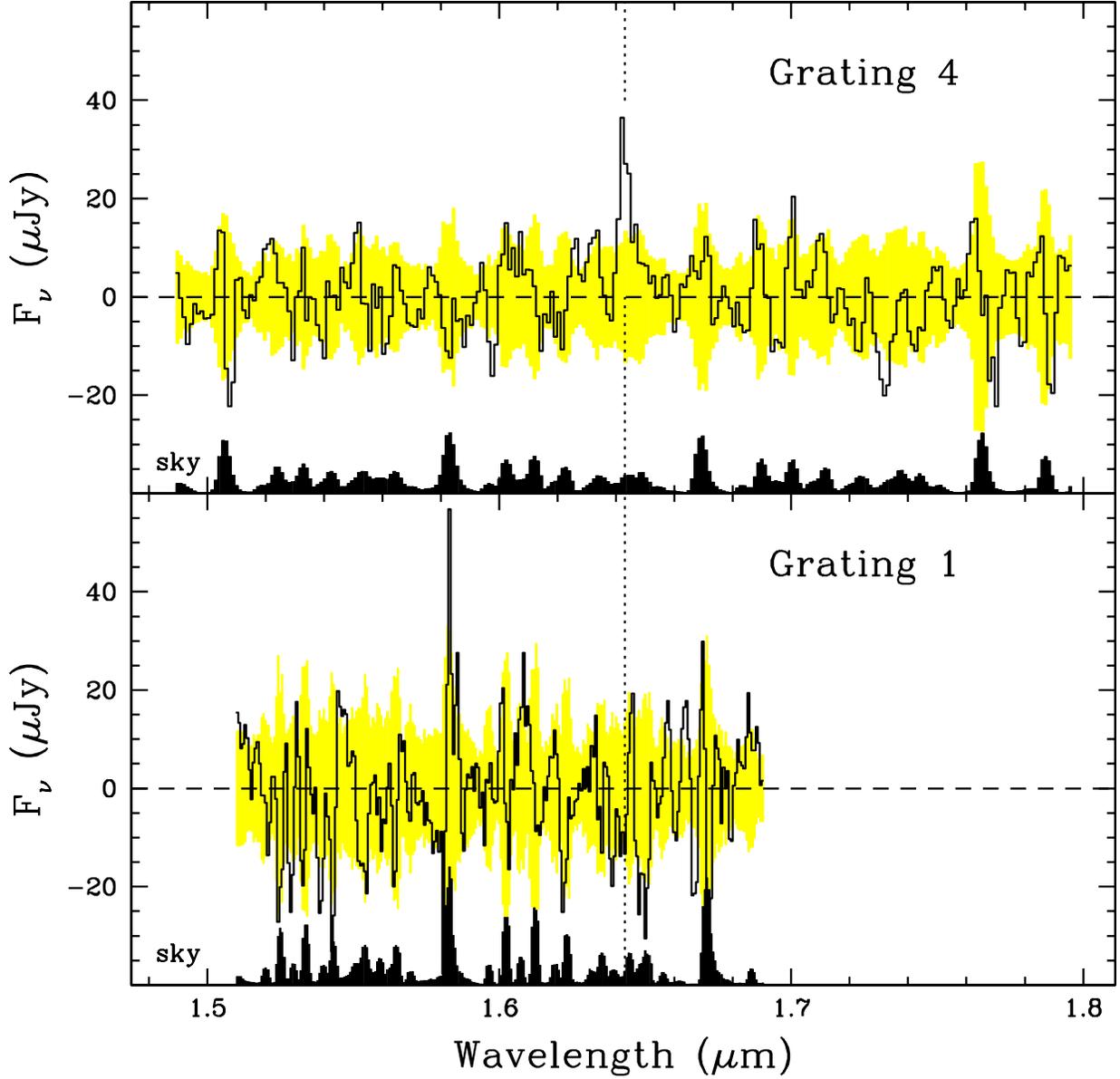}{6.3in}{0}{85}{85}{-250}{-120}
\caption{
%\figcaption[f03.ps]{
$H$--band spectrograms of \jdrop\ obtained with CRSP at the KPNO~4m.
The spectra have been smoothed by a 3--pixel boxcar.  The lightly
shaded regions show the $\pm 1\sigma$ noise level of the data (before 
smoothing), and greatly suppressed sky spectra are plotted along the bottom 
axis.  The low--resolution grating 4 spectrogram (top) shows a possible emission 
line at $\lambda 1.643\mu$m, but this is not detected in the subsequent,
higher resolution grating 1 spectrogram (bottom), leaving its reality
in doubt.
}
\end{figure}

\begin{figure}
\centerline{\hbox{
\psfig{figure=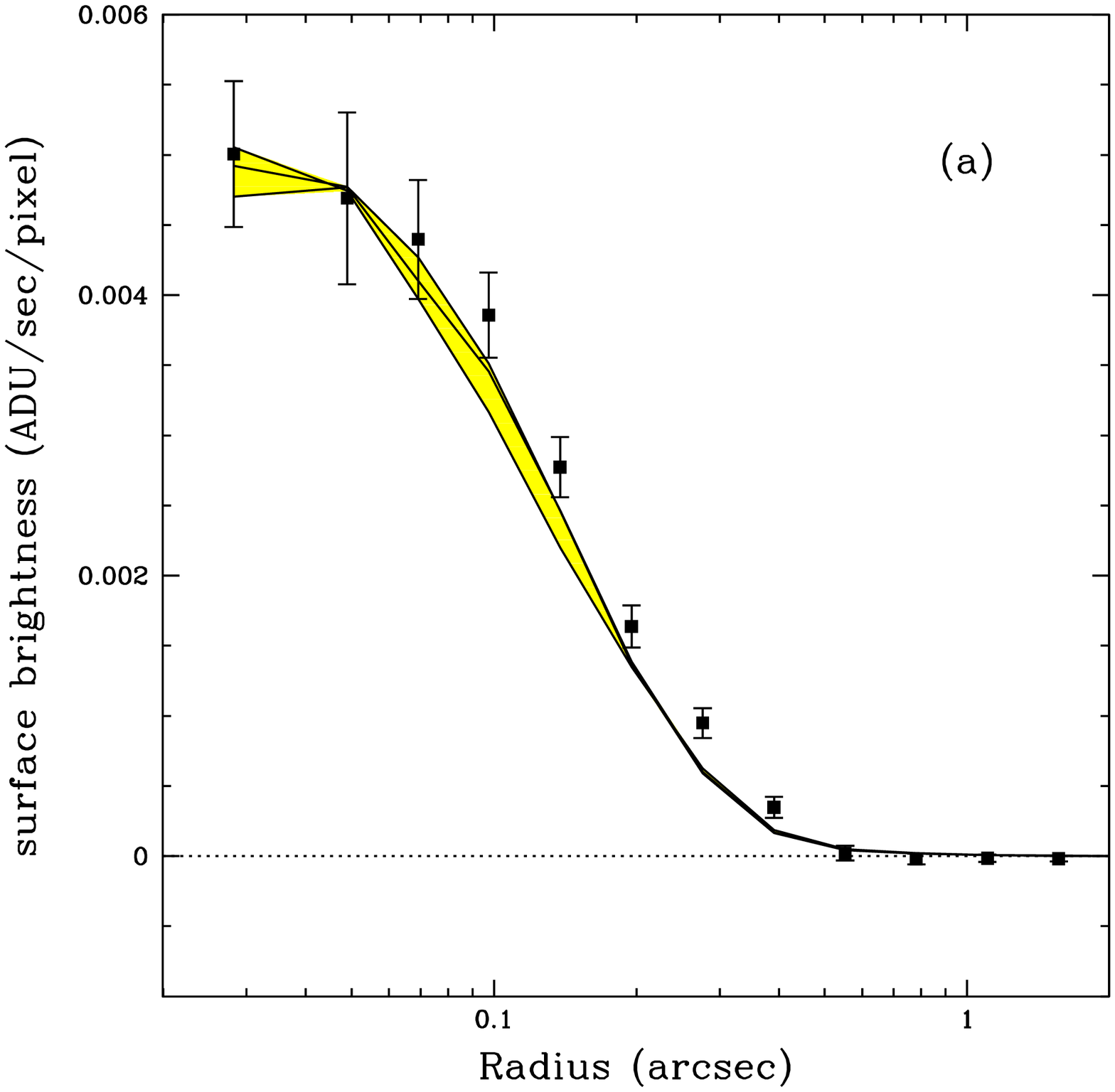,width=3.5in}
\psfig{figure=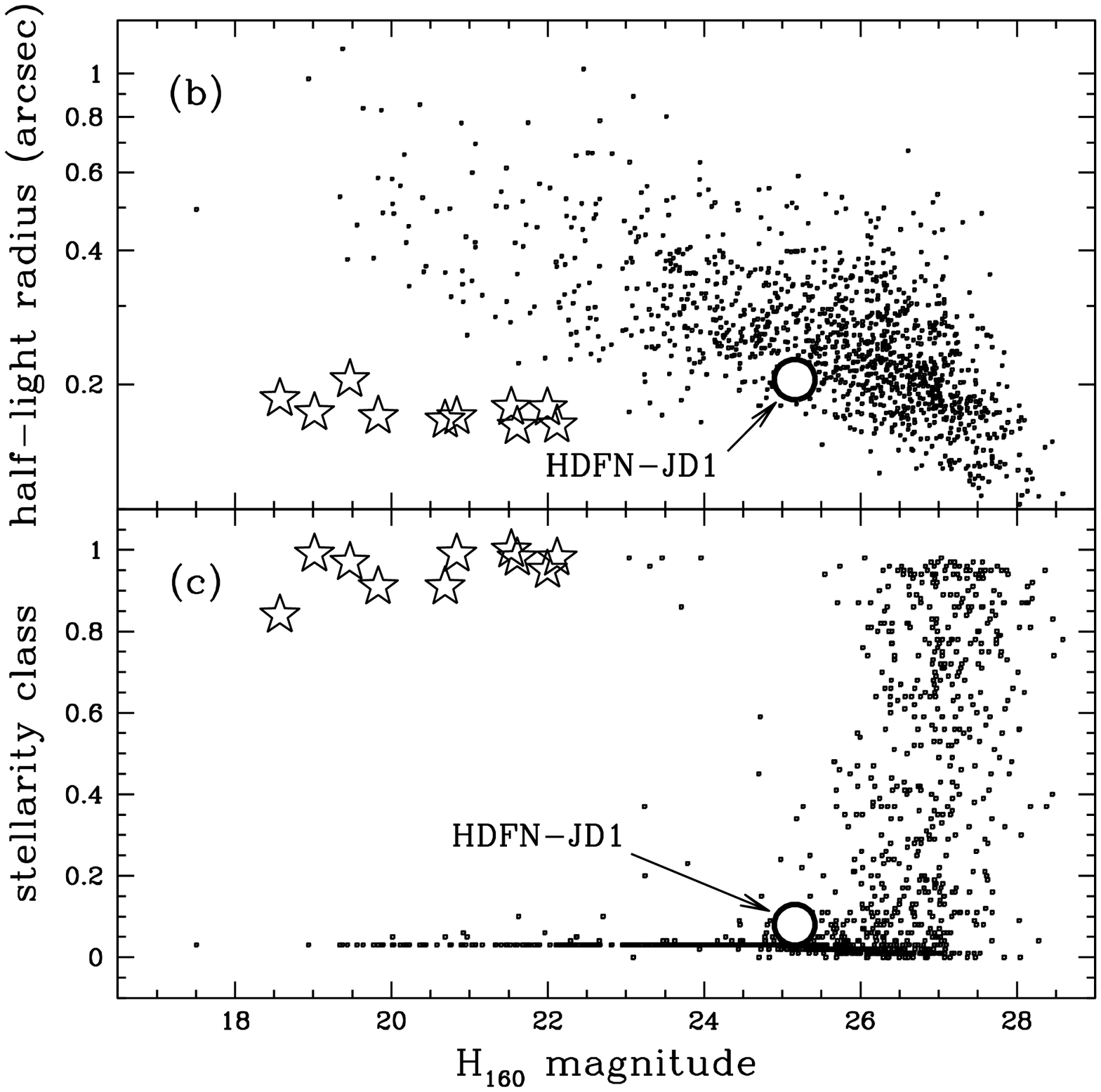,width=3.5in}
}}
\caption{
%\figcaption[f04a.ps,f04bc.ps]{
{\it Left:} (a) Radial surface brightness profile of \jdrop\ (points) compared to
three faint, isolated HDF stars (lines, shaded region) which have been registered
and scaled by non--negative cross--correlation.  {\it Right top:} (b) $\H160$ half 
light radius vs.\ magnitude for HDF objects.  Known stars are indicated by star 
symbols, and \jdrop\ is labeled.  {\it Right bottom:} (c) SExtractor stellar 
classifier (at \H160) for HDF objects, with symbols as in (b).  
}
\end{figure}

\begin{figure}
\plotfiddle{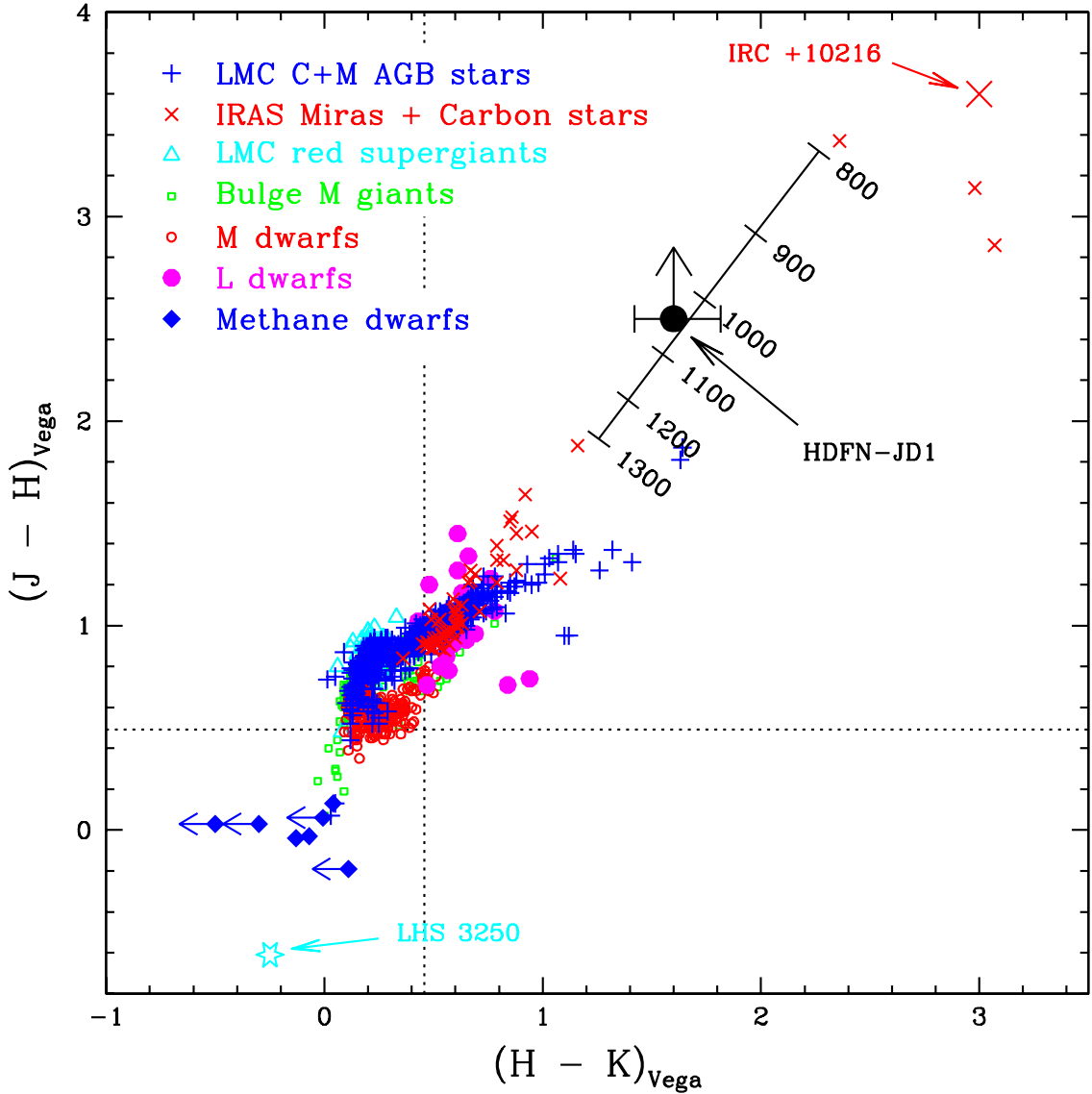}{5.5in}{0}{80}{80}{-250}{-130}
\caption{
%\figcaption[f05.ps]{
$J-H$ vs.\ $H-K$ color--color diagram (on the conventional, 
Vega--normalized system) comparing \jdrop\ to a variety of cool or 
reddened stars.  The NICMOS photometry has been converted 
to standard infrared magnitudes using synthetic color corrections,
and is plotted as a $2\sigma$ lower limit in $J-H$.  
The dotted lines mark the colors of a flat $f_{\nu}$ spectrum 
(i.e., zero AB colors).  The stellar data are taken from 
Frogel, Mould \& Blanco 1990 and Costa \& Frogel 1996 (LMC AGB stars), 
Whitelock \etal 1994, 1995 (IRAS Miras and carbon stars),
Oestreicher, Schmidt--Kaler, \& Wargau 1997 (supergiants),
Frogel \etal 1990 (M giants),
Leggett 1992 and Leggett \etal 1998 (M dwarfs),
Kirkpatrick \etal 1999 (L dwarfs), and
Burgasser \etal 1999, Leggett \etal 1999, and Strauss \etal 1999 (methane dwarfs).
IRC+10216 (Becklin \etal 1969) is also shown, along with LHS~3250 
(Harris \etal 1999) which is believed to be a very cool DC white dwarf.
The solid track shows the colors of a black body with labeled 
temperatures (in K).  
}
\end{figure}

\begin{figure}
\plotfiddle{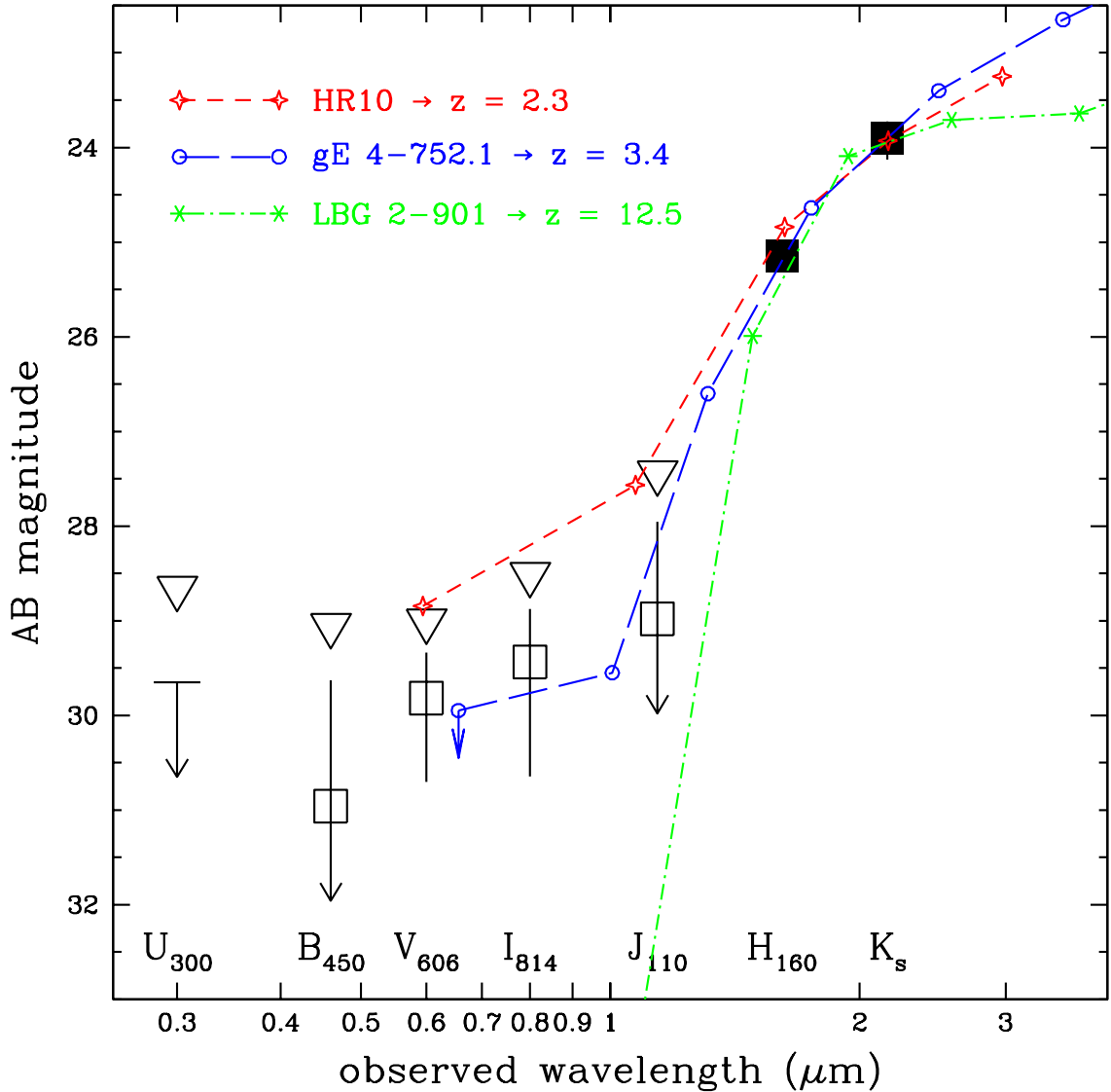}{5.5in}{0}{80}{80}{-250}{-130}
\caption{
%\figcaption[f06.ps]{
Spectral energy distribution (SED) of \jdrop.  Open squares mark
measurements with $S/N < 2$.  Error bars ($\pm 1\sigma$) are superimposed,
except when $S/N < 1$, where the downward error bar is shown as an arrow.
Triangles mark $2\sigma$ magnitude limits for \U300 through \J110.
Three SEDs of other, real galaxies, artificially redshifted and normalized
to the $K_s$ flux of \jdrop, are superimposed for comparison:  the dusty 
galaxy HR10 shifted to $z = 2.3$, the HDF giant elliptical galaxy 4-752.1 
shifted to $z = 3.4$, and the HDF Lyman break galaxy 2--901 shifted to 
$z = 12.5$.  For the latter, the flux point shortward of Lyman~$\alpha$ 
has been reduced to account for the assumed opacity increase of the 
Lyman~$\alpha$ forest at $z=12.5$, and flux shortward of the Lyman limit 
has been set to zero.  For HDF 4--752.1, the shortest wavelength measurement
(from the HDF \U300 image) probably overestimates the real UV flux from 
the gE galaxy because of a red leak in the WFPC2 F300W filter;  this is 
indicated by a downward--pointing arrow.  All three SEDs are qualitatively 
similar to that of \jdrop.
}
\end{figure}

\begin{figure}
\plotfiddle{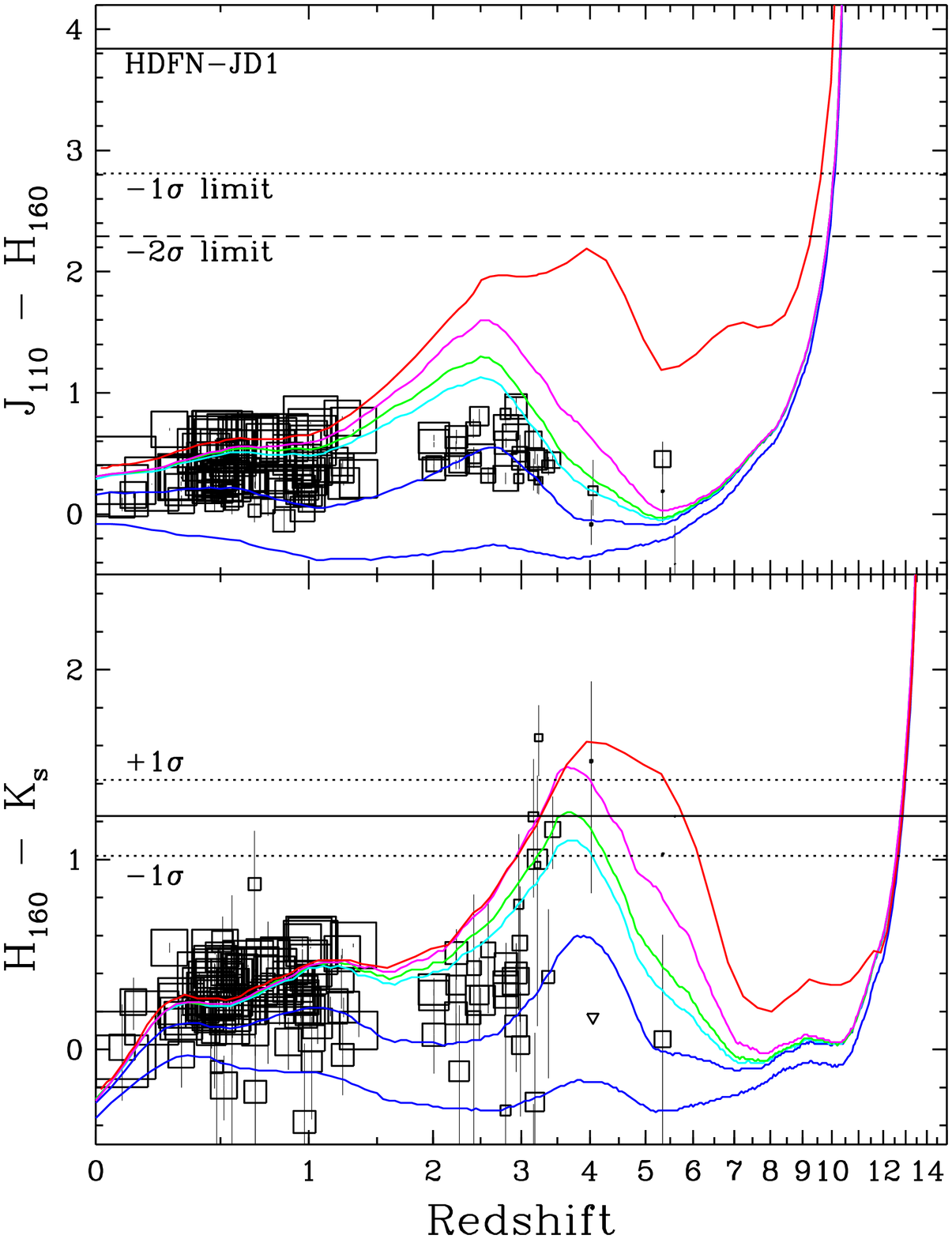}{6in}{0}{65}{65}{-200}{-20}
\caption{
%\figcaption[f07.ps]{
Infrared colors (on the AB system) versus redshift for galaxies
with spectroscopic redshifts in the HDF--N.  Point sizes scale with
the \H160 magnitudes.  The solid horizontal lines 
mark the nominal colors of \jdrop, and the dotted lines mark the 
$1\sigma$ uncertainties for $\H160 - K$ and lower color limit for
$\J110 - \H160$;  the dashed line for $\J110 - \H160$ marks the 
$2\sigma$ lower color limit.  The curves show the expected colors of 
various model galaxy types.  The reddest of these is a Bruzual \& Charlot 
(1996) model for an elliptical galaxy formed at $z = 15$ in a $10^8$ year 
burst and evolving passively in a cosmology where $\Omega_M = 0.3$, 
$\Omega_\Lambda = 0.7$, and $H_0 = 70$~km~s$^{-1}$~Mpc$^{-1}$.  
The other curves are unevolving models matching the colors of local
spirals and unreddened starburst galaxies.  The evolving elliptical 
model matches the $\H160 - K$ color of \jdrop\ at $3 \simlt z \simlt 6$, 
but never quite reaches the $2\sigma$ limit on $\J110 - \H160$.  
The HI opacity of the intergalactic medium has been included by 
extrapolating the models of Madau (1995) to $z = 15$.  At these large 
redshifts, the IGM transmission is effectively a step function 
at Lyman~$\alpha$.  Redshifts were compiled from Cohen \etal (1996), 
Steidel \etal (1996b), Lowenthal \etal (1997), Dickinson (1998), 
Hogg \etal (1998), Weymann \etal (1998) and Spinrad \etal (1998),
and Barger \etal (1999).
}
\end{figure}


\clearpage

\begin{references}

\reference{} Aussel, H., Cesarsky, C.J., Elbaz, D., \& Starck, J.L. 1999, 
	\aa, 342, 313

\reference{} Barger, A.J., Cowie, L.L., \& Sanders, D.B. 1999, \apj, 518, L5

\reference{} Barger, A.J., Cowie, L.L., Trentham, N., Fulton, E., 
	Hu, E.M., Songaila, A., \& Hall, D. 1999, \aj, 117, 102

\reference{} Becklin, E.E., Frogel, J.A., Hyland, A.R., Kristian, J., \& Neugebauer, G.
	1969, \apj, 158, L133

\reference{} Ben\'{\i}tez, N., Broadhurst, T., Bouwens, R., Silk, J., \& Rosati, P.
	1999, \apj, 515, L65

\reference{} Bertin, E., \& Arnouts, S. 1996, \aasup, 117, 393

\reference{} Brunner, R.J., Connolly, A.J., \& Szalay, A.S. 1999, \apj, 516, 563.

\reference{} Bruzual, G., \& Charlot, S.\ 1993, \apj, 405, 538
 
\reference{} Bruzual, G., \& Charlot, S.\ 1996, personal communication

\reference{} Budavari, T., \etal 1999, in preparation

% \reference{} Burgasser, A.J., Kirkpatrick, J.D., Brown, M.E., Reid, I.N.,
%	Gizis, J.E., Dahn, C.C., Monet, D.G., Beichman, C.A., Liebert, J.,
%	Cutri, R.M., \& Skrutskie, M.F. 1999, \apj, 522, L65
 
\reference{} Burgasser, A.J., \etal 1999, \apj, 522, L65
 
% \reference{} Burrows, A., Marley, M., Hubbard, W.B., Lunine, J.I., Guillot, T.,
%	Saumon, D., Freedman, R., Sudarsky, D., \& Sharp, C. 1997, \apj, 491, 856

\reference{} Burrows, A., \etal 1997, \apj, 491, 856

\reference{} Cimatti, A., Andreani, P., Rottgering, H., \& Tilanus, R. 1998,
	\nature, 392, 895

\reference{} Cohen, J.G., Cowie, L.L., Hogg, D.W., Songaila, A.,
        Blandford, R., Hu, E.M., \& Shopbell, P.  1996, \apj, 471, L5

\reference{} Costa, E., \& Frogel, J.A. 1996, \aj, 112, 2607

%\reference{} Cuby, J.G., Saracco, P., Moorwood, A.F.M., D'Odorico, S., Lidman, C.,
%	Comer\'on, F., \& Spyromilio, J. 1999, \aa, submitted

\reference{} D\'esert, F.-X., Puget, J.-L., Clements, D.L., P\'erault, M., 
	Abergel, A., Bernard, J.-P., \& Cesarsky, C.J. 1999, \aa, 342, 363

\reference{} Dey, A, Graham, J.R., Ivison, R.J., Smail, I., Wright, G.S.,
	\& Liu, M.C. 1999, \apj, 519, 610

\reference{} Dey, A. 1998, in {\it The Most Distant Radio Galaxies,} 
	proceedings of the 1997 KNAW colloquium, astro--ph/9802163

\reference{} Dickinson, M. 1997, in {\it Galaxy Scaling Relations,} eds. 
	L.N.\ da~Costa \& A.\ Renzini, (Berlin: Springer), p.~215

\reference{} Dickinson, M. 1998, in {\it The Hubble Deep Field,} eds. M.~Livio,
	S.M.~Fall \& P.~Madau, (Cambridge: Cambridge University Press), p.~219

\reference{} Dickinson, M. 1999, in {\it After the Dark Ages:  When Galaxies
	Were Young,} eds.\ S.S.\ Holt \& E.P.\ Smith, (Woodbury: AIP), p.~122

\reference{} Dickinson, M., \etal 1999, in preparation

\reference{} Downes, D., \etal 1999, \aa, 347, 809

\reference{} Dunlop, J., Peacock, J., Spinrad, H., Dey, A., Jiminez, R.,
	Stern, D., \& Windhorst, R. 1996, \nature, 381, 581

\reference{} Eke, V.R., Cole, S., \& Frenk, C.S. 1996, \mnras, 282, 263

\reference{} Elias, J.H., Frogel, J.A., Matthews, K., \& Neugebauer, G. 1982, 
	\aj, 87, 1029

\reference{} Elston, R., Rieke, G.H., \& Rieke, M.J. 1998, \apj, 331, L77

\reference{} Elston, R., Rieke, M.J., \& Rieke, G.H. 1999, \apj, 341, 80

\reference{} Fern\'andez--Soto, A., Lanzetta, K.M., \& Yahil, A. 1999,
	\apj, 513, 34

\reference{} Frogel, J.A., Mould, J., \& Blanco, V.M. 1990, \apj, 352, 96

\reference{} Frogel, J.A., Terndrup, D.M., Blanco, V.M., \& Whitford, A.E. 1990, 
	\apj, 353, 494

\reference{} Fruchter, A.S., and Hook, R.N. 1999, \pasp, submitted

\reference{} Gardner, J.P., Sharples, R.M., Frenk, C.S., \& Carrasco, B.E. 1997,
	\apj, 480, L99

\reference{} Gilliland, R.L., Nugent, P.E., \& Phillips, M.M. 1999, \apj, 521, 30

% \reference{} Goldschmidt, P., Oliver, S.J., Serjeant, S.B.G., Baker, A., 
% 	Eaton, N., Efstathiou, A., Grupponi, C., Mann, R.G., Mobasher, B., 
% 	Rowan--Robinson, M., Sumner, T.J., Danese, L.,
% 	Elbaz, D., Franceschini, A., Egami, E., Kontizas, M., Lawrence, A.,
% 	McMahon, R., Norgaard--Nielsen, H.U., Perez--Fournon, I., 
% 	\& Gonzalez--Serrano, J.I. 1997, \mnras, 289, 465

\reference{} Goldschmidt, P., \etal 1997, \mnras, 289, 465

\reference{} Graham, J.R., \& Dey, A. 1996, \apj, 471, 720

\reference{} Guhathakurta, P., Tyson, J.A., \& Majewski, S.R. 1990,
        \apj, 357, L9

\reference{} Haehnelt, M.G., Natarajan, P., \& Rees, M.J. 1998,
	\mnras, 300, 817

\reference{} Haiman, Z., \& Loeb, A. 1998, \apj, 503, 505

\reference{} Hansen, B.M.S. 1998, \nature, 394, 860

\reference{} Harris, H.C., Dahn, C.C, Vrba, F.J., Harden, A.A.,
	Liebert, J., Schmidt, G.D., \& Reid, I.N 1999, \apj, in press

\reference{} Hogg, D.W., et al. 1998, \aj, 115, 1418

\reference{} Hu, E.M., \& Ridgway, S.E. 1994, \aj, 107, 1303

% \reference{} Hughes, D., Serjeant, S., Dunlop, J., Rowan--Robinson, M., 
%	Blain, A., Mann, R., Ivison, R., Peacock, J., Efstathiou, A., Gear, W.,
%	Oliver, S.J., Lawrence, A., Longair, M., Goldschmidt, P., Jenness, T.
%	1998, \nature, 394, 241

\reference{} Hughes, D., \etal 1998, \nature, 394, 241

\reference{} Joyce, D. 1995, {\it Cryogenic Spectrograph User's Manual,} 
	(Tucson: NOAO)

\reference{} Kirkpatrick, \etal 1999, \apj, 519, 802

\reference{} Krist, J., \& Hook, R. 1997, {\it The Tiny Tim User's Guide} 
	(version 4.4), (Baltimore: STScI)

\reference{} Lanzetta, K.M., Yahil, A., \& Fern\'andez--Soto, A. 1996,
	\nature, 381, 759

\reference{} Lanzetta, K.M., Yahil, A., \& Fern\'andez--Soto, A. 1998,
	\aj, 116, 1066

\reference{} Lauer, T.R. 1999, \pasp, submitted

\reference{} Leggett, S.K. 1992, \apjs, 82, 351

\reference{} Leggett, S.K., Allard, F., \& Hauschildt, P.H. 1998, \apj, 509, 836

\reference{} Leggett, S.K., Toomey, D.W., Geballe, T.R., \& Brown, R.H. 1999,
	\apj, 517, L139

\reference{} Loveday, J., 1999, \mnras, in press

% \reference{} Lowenthal, J.D., Koo, D.C., Guzman, R., Gallego, J.,
%        Phillips, A.C., Faber, S.M., Vogt, N.P., Illingworth, G.D.,
%        \& Gronwall, C. 1997, \apj, 481, 673

\reference{} Lowenthal, J.D., \etal 1997, \apj, 481, 673

\reference{} Madau, P. 1995, \apj, 441, 18

\reference{} Maoz, D. 1997, \apj, 490, L135

\reference{} Matthews, K.Y., \& Soifer, B.T. 1994, in {\it IR Astronomy With 
Arrays,} ed.\ I.\ McLean, (Dordrecht: Kluwer), 239

\reference{} Mobasher, B., Sharples, R.M., \& Ellis, R. 1993, \mnras, 263, 560

\reference{} Morton, D.C., Spinrad, H., Bruzual, G., \& Kurucz, R.L. 1977,
	\apj, 212, 438

%\reference{} Nakajima, T., Oppenheimer, B.R., Kulkarni, S.R., Golimowski, D.A.,
%	Matthews, K., \& Durrance, S.T. 1995, \nature, 378, 463

\reference{} Oestreicher, M.O., Schmidt--Kaler, Th., \& Wargau, W. 1997,
	\mnras, 289, 729

\reference{} Reimers, D., Vogel, S., Hagen, H.-J., Engels, D., Groot, D.,
	Wamsteker, W., Clavel, J., \& Rosa, M.R. 1992, \nature, 360, 561

\reference{} Richards, E.A., Kellerman, K.I., Fomalont, E.B., Windhorst, R.A.,
	\& Partridge, R.B. 1998, \aj, 116, 1039

\reference{} Richards, E.A. 1999, \apj, 513, L9.

\reference{} Saumon, D., \& Jacobson, S.B. 1999, \apj, 511, L107

% \reference{} Skinner, C.J., Bergeron, L.E., Schultz, A.B., MacKenty, J.W.,
%	Storrs, A., Freudling, W., Axon, D., Bushouse, H., Calzetti, D., Colina, L.,
%	Daou, D., Gilmore, D., Holfeltz, S.T., Najita, J., Noll, K., Ritchie, C., 
%	Sparks, W.B., \& Suchkov, A. 1998, SPIE, 3354, 2

\reference{} Skinner, C.J., \etal 1998, SPIE, 3354, 2

\reference{} Smail, I., Ivison, R.J., Kneib, J.-P., Cowie, L.L., Blain, A.W.,
	Barger, A.J., Owen, F.N., \& Morrison, G.E. 1999, \mnras, in press

\reference{} Spinrad, H., Dey, A., Stern, D., Dunlop, J., Peacock, J.,
	Jimenez, R., \& Windhorst, R. 1997, \apj, 484, 581

\reference{} Spinrad, H., Stern, D., Bunker, A., Dey, A., Lanzetta, K.M., 
	Yahil, A., Pascarelle, S., \& Fern\'andez--Soto, A. 1998, \aj, 116, 2617

\reference{} Steidel, C.C., Giavalisco, M., Pettini, M., Dickinson, M.,
        \& Adelberger, K.L. 1996a, \apj, 462, L17
 
\reference{} Steidel, C.C., Giavalisco, M., Dickinson, M., \& Adelberger, K.L.
        1996b, \aj, 112, 352

\reference{} Steidel, C.C., Adelberger, K.L., Giavalisco, M., Dickinson, M., 
	\& Pettini, M. 1999, \apj, 519, 1

% \reference{} Stiavelli, M., Treu, T., Carollo, C.M., Rosati, P., Viezzer, R., 
%	Casertano, S., Dickinson, M., Ferguson, H., Fruchter, A., Madau, P., 
%	Martin, C., \& Teplitz, H. 1999, \aa, 343, L25

\reference{} Stiavelli, M., \etal 1999, \aa, 343, L25

% \reference{} Strauss, M., Fan, X., Gunn, J.E., Leggett, S.K., Geballe, T.R.,
% 	Pier, J.R., Lupton, R.H., Knapp, G.R., Annis, J., Brinkmann, J.,
% 	Crocker, J.H., Csabai, I., Golimowski, D.A., Harris, F.H., Hennessy, G.S.,
% 	Hindsley, R.B., Ivezi\'c, Z., Lamb, D.Q., Munn, J.A., Newberg, H.J., 
% 	Rechenmacher, R., Schneider, D.P., Smith, J.A., Stoughton, C.,
% 	Tucker, D.L., Waddell, P., \& York, D.C. 1999, \apj, submitted

\reference{} Strauss, M., \etal 1999, \apj, submitted

\reference{} Szalay, A.S., Connolly, A.J., \& Szokoly, G.P. 1999, \aj, 117, 68

\reference{} Szokoly, G.P., Subbarao, M.U., Connolly, A.J., \& Mobasher, B. 1998, 
	\apj, 492, 452

\reference{} Thompson, D., \etal 1999a, \apj, in press

\reference{} Thompson, R.I., Rieke, M., Schneider, G., Hines, D.C., 
	\& Corbin, M.R. 1998, \apj, 492, L95

\reference{} Thompson, R.I., Storrie--Lombardi, L.J., Weymann, R.J., Rieke, M., 
	Schneider, G., Stobie, E., \& Lytle, D. 1999b, \aj, 117, 17

% \reference{} Tinney, C.G., Mould, J.R., \& Reid, I.N. 1993, \aj, 105, 1045

\reference{} Veron, P. 1983, in {\it Quasars and Gravitational Lenses,}
	Universit\'e de Li\`ege, Institute d'Astrophysique, p.~210

\reference{} Waddington, I., Windhorst, R. A., Cohen, S. H., Partridge, R. B.,  
	Spinrad, H., \& Stern, D. 1999, ApJL, submitted

\reference{} Warren, S.J., Hewitt, P.C., Irwin, M.J., McMahon, R.G.,
        Bridgeland, M.T., Bunclark, P.S., \& Kibblewhite, E.J. 1987,
        Nature, 325, 131

\reference{} Weymann, R.J., Stern, D., Bunker, A., Spinrad, H., Chaffee, F.H.,
	Thompson, R.I., \& Storrie--Lombardi, L.J. 1998, \apj, 505, L95

\reference{} Whitelock, P., Menzies, J., Feast, M., Marang, F., 
	Carter, B., Roberts, G., Catchpole, R., \& Chapman, J.
	1994, \mnras, 267, 711

\reference{} Whitelock, P., Menzies, J., Feast, M., Catchpole, R., Marang, F.,
	\& Carter, B. 1995, \mnras, 276, 219

% \reference{} Williams, R.E., Blacker, B., Dickinson, M., Dixon, W.V.D.,
%        Ferguson, H.C., Fruchter, A.S., Giavalisco, M.,
%        Gilliland R.L., Heyer, I., Katsanis, R., Levay, Z., Lucas, R.A.,
%        McElroy, D.B., Petro, L., Postman, M., Adorf, H.-M., Hook, R.N.
%        1996, \aj, 112, 1335

\reference{} Williams, R.E., \etal 1996, \aj, 112, 1335

\reference{} Williams, R.E., \etal 1999, in preparation

\end{references}
\end{document}